\documentclass[a4paper,11pt]{article}
\pdfoutput=1 % if your are submitting a pdflatex (i.e. if you have
\usepackage{jheppub} % for details on the use of the package, please\eta
\usepackage{amssymb} 
\usepackage{amsmath}
\usepackage{mathtools}
\usepackage{amsfonts}    
\usepackage{dsfont}
\usepackage{pdfpages}
\usepackage{verbatim}
\usepackage{tensor}
\usepackage{mathrsfs}
\usepackage[mathscr]{euscript}
\usepackage{tikz}
\usepackage{braket}

\let\oldbibliography\thebibliography
\renewcommand{\thebibliography}[1]{\oldbibliography{#1}
\setlength{\itemsep}{4.14pt}} %Reducing spacing in the bibliography.

\allowdisplaybreaks
\usepackage{mathrsfs}
\usepackage[mathscr]{euscript}

\numberwithin{equation}{section}

\def\be{\begin{equation}}
\def\ee{\end{equation}}

\title{\boldmath Free field world-sheet correlators for ${\rm AdS}_3$}

\author[a]{Andrea Dei}
\author[a]{\!\!, Matthias R.~Gaberdiel}
\author[b]{\!\!, Rajesh Gopakumar}
\author[a]{and Bob Knighton}

\affiliation[a]{Institut f\"ur Theoretische Physik, ETH Z\"urich \\
\hspace*{0.3cm} Wolfgang-Pauli-Stra{\ss}e 27, 8093 Z\"urich, Switzerland}
\affiliation[b]{International Centre for Theoretical Sciences-TIFR, \\
\hspace*{0.3cm} Shivakote, Hesaraghatta Hobli, Bengaluru North, India 560 089}

\emailAdd{adei@itp.phys.ethz.ch}
\emailAdd{gaberdiel@itp.phys.ethz.ch}
\emailAdd{rajesh.gopakumar@icts.res.in}
\emailAdd{robejr@ethz.ch}

\abstract{We employ the free field realisation of the $\mathfrak{psu}(1,1|2)_1$ world-sheet theory to constrain the correlators of string theory on ${\rm AdS}_3\times {\rm S}^3\times \mathbb{T}^4$ with unit NS-NS flux. In particular, we directly obtain the unusual delta function localisation of these correlators onto branched covers of the boundary ${\rm S}^2$ by the (genus zero) world-sheet --- this is the key property which makes the equivalence to the dual symmetric orbifold manifest. In our approach, this feature follows from a remarkable `incidence relation' obeyed by the correlators, which is reminiscent of a twistorial string description. We also illustrate our results with explicit computations in various special cases.}

\begin{document}
\maketitle
\flushbottom

\section{Introduction}

Deriving the ${\rm AdS}/{\rm CFT}$ correspondence, even in a handful of cases, is likely to teach us a lot about the inner workings of this holographic duality. Recently, progress in this direction has been made in a special case which relates the small radius or tensionless limit of ${\rm AdS}_3\times {\rm S}^3\times \mathbb{T}^4$ (with $k=1$ units of NS-NS flux) to the free symmetric product orbifold CFT,
${\rm Sym}^N (\mathbb{T}^4)$. The first piece of evidence for this duality was the agreement of the full spectrum in the large $N$ limit \cite{Eberhardt:2018ouy}, see also \cite{Gaberdiel:2018rqv}. It was subsequently shown in \cite{Eberhardt:2019ywk} how the correlators of the two descriptions  agree manifestly. 
More specifically, it was shown that the world-sheet correlator on the left-hand-side of 
\begin{multline} \label{eq:corresp}
\int_{{\cal M}_{g,n}} \!\!\! \! \!\! \big\langle \mathcal{V}^{w_1}_{h_1}(x_1; z_1)\mathcal{V}^{w_2}_{h_2}(x_2; z_2) \ldots \mathcal{V}_{h_n}^{w_n}(x_n; z_n) \big\rangle_{\Sigma_{g,n}}  \\
= \big\langle {\cal O}^{(w_1)}_{h_1}(x_1){\cal O}^{(w_2)}_{h_2}(x_2) \ldots {\cal O}^{(w_n)}_{h_n}(x_n) \big\rangle_{\text{S}^2} \Big|_{g} \ ,
\end{multline}
becomes the orbifold CFT correlator on the right hand side by virtue of the fact that it localises to the points in the moduli space ${\cal M}_{g,n}$ where a holomorphic covering map from the genus $g$ world-sheet to the boundary sphere exists. This then reproduces manifestly, i.e.\ without any explicit computation, the symmetric orbifold correlator since the latter is determined in terms of these covering maps \cite{Lunin:2001,Pakman:2009}.

The strategy adopted in \cite{Eberhardt:2019ywk} for genus zero (generalised to higher genus in 
\cite{Eberhardt:2020akk}), to show this striking localisation, was to use the Ward identities for the $\mathfrak{sl}(2, \mathbb{R})_k$ currents on the world-sheet. This required understanding how to deduce constraints for correlators of $w$-spectrally flowed $\mathfrak{sl}(2, \mathbb{R})_k$ representations, which was the main technical advance of  \cite{Eberhardt:2019ywk}. A special solution to these constraints had the property that it was determined in terms of a holomorphic covering map of the boundary ${\rm S}^2$ by the world-sheet, $x=\Gamma(z)$, where $\Gamma$ maps the world-sheet insertion points $z_i$ to the spacetime insertion points $x_i$, $x_i=\Gamma(z_i)$, and $z_i$ is a branch-point of ramification index $w_i$. 

Covering maps characterised by these branching data $\{ z_i, w_i \}$ are essentially fixed if three of the image points $x_i$ are prescribed, i.e.\ the remaining $(n-3)$ points $x_i$ are determined in terms of the branched cover (up to finitely many choices). Conversely, in terms of the $z_i$, this means that for fixed $\{x_i,w_i\}$, the covering map only exists for isolated combinations of the $z_i$. In fact, the special solution of \cite{Eberhardt:2019ywk} had delta function support in $z_i$ on the co-dimension $(n-3)$ locus on the moduli space ${\cal M}_{0,n}$ where this is the case. This is precisely what is needed to reproduce the orbifold CFT correlators on the right hand side of 
\eqref{eq:corresp} as per the Lunin-Mathur construction \cite{Lunin:2001, Pakman:2009}. In a second strand of argument, it was shown in \cite{Eberhardt:2019ywk} that there exists a set of semi-classically exact solutions of the ${\rm AdS}_3$ sigma model whose contributions reproduce the correct weight factors for each of these branched covers. Taken together, these two arguments gave strong evidence that the correlators on both sides of \eqref{eq:corresp} are manifestly equal. 
\smallskip

In this paper, we tighten the first argument from \cite{Eberhardt:2019ywk}: while \cite{Eberhardt:2019ywk} showed that \emph{a} solution to the Ward identities with these properties exists, it was not clear  that this is the only solution, although there was good circumstantial evidence that it is the physically relevant one.\footnote{In \cite{Hikida:2020kil} the same correlators were subsequently analysed using the Wakimoto free field realisation of $\mathfrak{sl}(2,\mathds{R})_{k+2}$. While the results of \cite{Hikida:2020kil} are compatible with \cite{Eberhardt:2019ywk}, this approach also did not tie down the answer further.}
 Furthermore, the analysis of \cite{Eberhardt:2019ywk} was essentially done in the NS-R framework, using only the bosonic $\mathfrak{sl}(2,\mathds{R})_{k+2}$ symmetry as in \cite{Maldacena:2000hw,Maldacena:2000kv,Maldacena:2001km}. While this is legitimate, it is a bit unsatisfactory since for $k=1$ the NS-R approach breaks down, and we should really be using the hybrid formulation as advocated in \cite{Eberhardt:2018ouy}.  Here we will start directly with the hybrid world-sheet formalism for the ${\rm AdS}_3$ string theory, which is described by a 
$\mathfrak{psu}(1,1|2)_k$ WZW model in the case of pure NS-NS flux \cite{Berkovits:1999im}. When the level $k=1$, there exists a free field realisation of this WZW model in terms of a pair of spin $\frac{1}{2}$ (symplectic) bosons ($\xi^{\pm}$), and a pair of spin $\frac{1}{2}$ fermions ($\psi^{\pm}$), together with their canonically conjugate fields. This is the same free field realisation which was used to deduce the string spectrum in  \cite{Eberhardt:2018ouy}. 

After describing the physical correlators in this language, taking into account subtleties of picture changing and charge conservation, we examine the constraints coming from the OPE of the symplectic bosons with the spectrally flowed vertex operators. It turns out that these can be nicely encapsulated in an incidence relation 
\begin{equation}
\Big{\langle} \bigl(\xi^-(z)+\Gamma(z)\, \xi^+(z)\bigr)\Big{\rangle}_{\rm phys} =0 \ .
\label{eq:incid}
\end{equation}
The brackets here denote the expectation value of the vertex operators $V^{w_i}(x_i;z_i)$, see eq.~\eqref{eq:magic}, and 
$\Gamma(z)$ is the covering map mentioned above with branching behaviour $w_i$ at the $z_i$ and 
$x_i=\Gamma(z_i)$ for $i=1,\ldots, n$. We show that this `incidence relation' implies that the correlators are  delta function localised to the locus where a covering map exists; it also fixes some of the additional dependence of the correlators, see eq.~\eqref{general-correlator}. The argument for this incidence relation is quite abstract (and general), but we also confirm it (and its consequences) explicitly in many examples. 

Since $x(z)=\Gamma(z)$ when the covering map exists,  the relation $\eqref{eq:incid}$ is very suggestive of an underlying twistor string like description for our system. In this language the $\xi^{\pm}$ play the role of twistor-like projective coordinates for the boundary $\mathbb{P}^1$. As we discuss in Section~\ref{twstr}, the action describing the free field variables is, in fact, a lower dimensional analogue of the twistor string action of Berkovits \cite{Berkovits:2004hg}. We believe this twistor like description holds the key to many of the underlying topological string features of the $k=1$ theory \cite{GG}.  Note that this topological behaviour is quite specific to the case with $k=1$, and appears to be different than the usual topological sector of string theory on ${\rm AdS}_3$ that was recently discussed in \cite{Costello:2020jbh,Li:2020nei}. Incidentally, the fact that the ${\rm AdS}_3$ world-sheet theory for $k=1$ is special was first noticed in a somewhat different context in \cite{Giveon:2005mi}, see also \cite{Giribet:2018ada} for subsequent developments.
\smallskip

We should mention that our analysis only concerns the chiral (and anti-chiral) correlators. In order to fully determine the correlators of the world-sheet theory (and obtain their OPE coefficients, etc.), we also have to solve the conformal bootstrap equations for $\mathfrak{psu}(1,1|2)_1$, which has, to our knowledge, not yet been done.\footnote{For the mini-superspace limit of ${\rm AdS}_3$ this was discussed in \cite{Ribault:2009ui}, using results for $H_3^+$ from \cite{Teschner:1997fv,Teschner:2001gi}, but this analysis does not include the spectrally flowed sectors. Most treatments of the spectrally flowed sectors on the other hand, see for example \cite{Ribault:2005ms,Giribet:2005mc,Baron:2008qf}, seem to set the spacetime variable $x$ uniformly to $x=0$ (or $x=\infty$), which is not appropriate in our context.} Given the results of this paper, together with the expected form of the general answer (see eq.~\eqref{full-corr}), this bootstrap programme should now be within reach. We should also mention that in order to determine the full string theory amplitudes, one also needs to include correctly the ghost contributions, the fermions, as well as the degrees of freedom coming from the $\mathbb{T}^4$. We have not yet attempted to do so in detail, although this should not be too difficult.

\medskip

The paper is organised as follows. We begin in Section~\ref{sec:WZW} with a review of the underlying WZW model and its free field realisation. In Section~\ref{sec:picturechanging} we explain some of the subtleties that arise for the correlators in the hybrid formalism; in particular, we explain how picture changing needs to be taken into account. Section~\ref{sec:twistor}  is the core of the paper: we demonstrate that the correlators satisfy the incidence relation, see eq.~(\ref{eq:magic}), and deduce various consequences from it, in particular, the fusion rules (Section~\ref{sec:fusion}), as well as the delta-function localisation property, see eq.~(\ref{corrstructfinal}). We also comment on the relation to the twistor string theory in Section~\ref{twstr}. We exemplify these findings with explicit calculations in Section~\ref{sec:examples}. In particular, we show how the symplectic boson Ward identities can be used to calculate these correlators directly, see Section~\ref{sec:Ward}. Section~\ref{sec:conclusions} contains our conclusions and outlook for future work. There are a number of Appendices where some of the more technical material has been collected. In particular, we give some details about the spectrally flowed representations in Appendix~\ref{app:spectral-flow}, and explain how to construct the covering map in Appendix~\ref{app:cover}. We also use the incidence relation in Appendix~\ref{app:explicit} to solve for the correlators, and demonstrate that some of the other constraint equations (that are not directly needed for the determination of the correlators) are also satisfied, see Appendix~\ref{app:eta1111}.  

\section{The world-sheet WZW model}\label{sec:WZW}

In this section we describe the free field realisation of our world-sheet theory in detail and fix our conventions. The key component we shall be focussing on is the $\mathfrak{psu}(1,1|2)$ WZW model at level $k=1$. Following \cite{Eberhardt:2018ouy}, see also \cite{Lesage:2002ch,Gotz:2006qp,Ridout:2010jk,Quella:2013oda}, it has a free field realisation 
in terms of two pairs of complex fermions and symplectic bosons, with (anti)-commutation relations\footnote{Relative to \cite{Eberhardt:2018ouy} we have renamed $\bar{\xi}^\alpha \mapsto \xi^\alpha$ and $\xi^\alpha \mapsto \eta^\alpha$, and similarly for the fermions, 
$\bar\psi^\alpha \mapsto \psi^\alpha$ and $\psi^\alpha\mapsto \chi^\alpha$. Note that all of these fields are left-moving fields, and they are not complex conjugates of one another; thus a notation without barred variables seems more appropriate.}
\begin{align}
\{\psi^\alpha_r,\chi^\beta_s\}=\epsilon^{\alpha\beta} \delta_{r, -s}\ , \qquad [\xi^\alpha_r,\eta^\beta_s]=\epsilon^{\alpha\beta} \delta_{r,-s}\ .
\end{align}
Here $\alpha,\beta\in\{\pm\}$, and we choose the convention that $\epsilon^{+-}=1 = - \epsilon^{-+}$. These free fields generate the algebra $\mathfrak{u}(1,1|2)_1$, where 
\begin{subequations}
\begin{align}
J^3_m&=-\tfrac{1}{2} (\eta^+\xi^-)_m-\tfrac{1}{2} (\eta^-\xi^+)_m\ , & K^3_m&=-\tfrac{1}{2} (\chi^+\psi^-)_m-\tfrac{1}{2} (\chi^-\psi^+)_m\ , \\
J^\pm_m&=(\eta^\pm\xi^\pm)_m\ , & K^\pm_m&=\pm(\chi^\pm\psi^\pm)_m\ , \\
S_m^{\alpha\beta+}&=(\chi^\beta \xi^\alpha)_m\ , & S_m^{\alpha\beta-}&=-(\eta^\alpha\psi^\beta)_m \ ,  \\
U_m&=-\tfrac{1}{2} (\eta^+\xi^-)_m+\tfrac{1}{2} (\eta^-\xi^+)_m\ , & V_m&=-\tfrac{1}{2} (\chi^+\psi^-)_m+\tfrac{1}{2} (\chi^-\psi^+)_m\ .
\end{align}\label{eq:u112-algebra}
\end{subequations}
The generators $J^a_m$ and $K^a_m$ describe $\mathfrak{sl}(2,\mathds{R})_1$ and $\mathfrak{su}(2)_1$, respectively, while the $S_m^{\alpha\beta\pm}$ generators are the supercharges. Their anti-commutator takes the form 
\begin{multline} 
\{S^{\alpha\beta\gamma}_m,S^{\mu\nu\rho}_n\}=km \epsilon^{\alpha\mu}\epsilon^{\beta\nu}\epsilon^{\gamma\rho}\delta_{m+n,0}-\epsilon^{\beta\nu}\epsilon^{\gamma\rho} c_a\tensor{\sigma}{_a^{\alpha\mu}} J^a_{m+n}\\
+\epsilon^{\alpha\mu}\epsilon^{\gamma\rho} \tensor{\sigma}{_a^{\beta\nu}} K^a_{m+n}+\epsilon^{\alpha\mu}\epsilon^{\beta\nu}\delta^{\gamma,-\rho} Z_{m+n}\ , \label{anticomm}
\end{multline}
where we raise and lower indices using the standard $\mathfrak{su}(2)$-invariant form
\be 
g_{+-}=g_{-+}=\tfrac{1}{2}\ , \qquad g_{33}=1\ , 
\ee 
and we have introduced the combinations 
\be 
Z_m=U_m+V_m\qquad\text{and}\qquad Y_m=U_m-V_m\ . 
\ee
Furthermore, the constant $c_a$ equals $-1$ for $a=-$, and $+1$ otherwise, and the $\sigma$-matrices are explicitly given by 
\begin{align}
\tensor{(\sigma^-)}{^+_-}&=2\ , & \tensor{(\sigma^3)}{^-_-}&=-1\ , & \tensor{(\sigma^3)}{^+_+}&=1\ , & \tensor{(\sigma^+)}{^-_+}&=2\ , \\
\tensor{(\sigma_-)}{^{--}}&=1\ , & \tensor{(\sigma_3)}{^{-+}}&=1\ , & \tensor{(\sigma_3)}{^{+-}}&=1\ , & \tensor{(\sigma_+)}{^{++}}&=-1\ ,
\end{align}
while all the other components vanish. Finally, $U_m$ and $V_m$ define two $\mathfrak{u}(1)$ algebras, with commutation relations 
\be
[U_m,U_n]  = - \frac{m}{2} \, \delta_{m,-n}  \ , \qquad 
[V_m,V_n]  = + \frac{m}{2} \, \delta_{m,-n}  \ . 
\ee
and hence $Y_m$ and $Z_m$ satisfy
\be\label{ZZ}
[Z_m,Z_n] = 0  = [Y_m,Y_n] \ ,  \qquad [Z_m,Y_n] = - m \, \delta_{m,-n} \ . 
\ee
Except for the non-trivial commutator with $Y_n$, $Z_m$ is central, and in order to obtain 
$\mathfrak{psu}(1,1|2)_1$ from the above algebra we need to set $Z_n=0$. More precisely this means that we take the coset by the $\mathfrak{u}(1)$ algebra generated by $Z_n$, i.e.\ we concentrate on the subspace of states that are annihilated by $Z_n$ with $n\geq 0$. We then consider the quotient space of this subspace where we divide out by the $Z_{-n}$ descendants with $n>0$ (which are null).

\subsection{The highest weight representations} 

At level $k=1$, $\mathfrak{psu}(1,1|2)_1$ has apart from the vacuum representation only one (highest weight) representation \cite{Eberhardt:2018ouy}, and this can also be understood in the above free field realisation. The vacuum representation arises from the NS-sector where both the symplectic bosons and the free fermions are half-integer moded,\footnote{Since the supercurrent generators involve one fermionic generator and one symplectic boson generator, the moding of all generators has to be the same in order for the $\mathfrak{psu}(1,1|2)$ generators to be integer moded.} and it is generated from a ground state satisfying 
\be
\eta^\pm_r \, |0\rangle = \xi^\pm_r \, |0\rangle = \chi^\pm_r \, |0\rangle = \psi^\pm_r \, |0\rangle = 0 \qquad r \geq \tfrac{1}{2} \ . 
\ee
Note that this state has the property that also 
\be
U_0 \, |0\rangle = V_0 \, |0\rangle = Z_0\,  |0\rangle = 0 \ . 
\ee
It is easy to see that it leads to the vacuum representation of $\mathfrak{psu}(1,1|2)$. 
\medskip

As regards the R-sector, let us first consider the symplectic boson generators. On the highest weight states, we let the symplectic boson zero modes act as 
\be\label{xizero}
\begin{array}{rclrcl}
\xi^+_0 |m_1,m_2\rangle & = &  \, |m_1,m_2+\tfrac{1}{2}\rangle \ , \qquad 
&
\eta^+_0 |m_1,m_2\rangle & = & 2 \, m_1 \, |m_1+\tfrac{1}{2},m_2\rangle \  , \\[3pt]
\xi^-_0 |m_1,m_2\rangle & = & - \, |m_1-\tfrac{1}{2},m_2\rangle \ , \qquad 
& {\eta}^-_0 |m_1,m_2\rangle & = & - 2 \, m_2 \, |m_1,m_2-\tfrac{1}{2}\rangle \ ,
\end{array}
\ee
where $m_1,m_2$ label the different highest weight states. In this convention we then have 
\begin{eqnarray} \label{eigenvs}
J^3_0 \, |m_1,m_2\rangle & =& (m_1+m_2) \, |m_1,m_2\rangle \label{J30} \\
U_0 \, |m_1,m_2\rangle & =& (m_1-m_2 - \tfrac{1}{2}) \, |m_1,m_2\rangle  \ . \label{U0}
\end{eqnarray}
The $\mathfrak{sl}(2,\mathds{R})$ Casimir on the highest weight states labelled by $|m_1,m_2\rangle$ equals 
\be
C^{\mathfrak{sl}(2,\mathds{R})} =  \frac{1}{4} - U_0^2 \ . 
\ee
Writing $C^{\mathfrak{sl}(2,\mathds{R})}  = - j (j-1)$, this then leads to the equation 
\be
j = m_1 - m_2 \ . 
\ee
As regards the fermionic generators, each highest weight state $|m_1,m_2\rangle$ leads to a 
Clifford representation with respect to the fermionic zero modes $\chi^\pm_0$ and $\psi^\pm_0$. We shall choose the convention that $|m_1,m_2\rangle$  is annihilated by the $+$-modes, 
\be
\chi^+_0 |m_1,m_2\rangle = \psi^+_0 |m_1,m_2\rangle= 0  \ . 
\ee
Then the action of the creation operators $\chi^-_0$ and $\psi^-_0$ leads to a $4$-dimensional space of states; with respect to $\mathfrak{su}(2)_1$, it decomposes into a doublet spanned by 
\be
{\bf 2}: \qquad  |m_1,m_2\rangle \ , \qquad  \hbox{and} \qquad 
 \chi^-_0 \psi^-_0 \, |m_1,m_2\rangle \ , 
\ee
as well as two singlet states 
\be
2\cdot {\bf 1}: \qquad 
 \chi^-_0  \, |m_1,m_2\rangle \ , \qquad \hbox{and} \qquad 
  \psi^-_0  \, |m_1,m_2\rangle  \ . 
\ee
In our conventions 
\be\label{V0eval}
V_0 = \left\{ \begin{array}{cl}  0 & \hbox{for the doublet} \\[2pt]
\pm \tfrac{1}{2} & \hbox{for the singlets.}
\end{array} \right. 
\ee
Thus the condition that $Z_0 = U_0 + V_0 = 0$ implies  that the possible ground states have to sit in the representations of $\mathfrak{sl}(2,\mathds{R}) \oplus \mathfrak{su}(2)$ 
\be
\bigl[ j=\tfrac{1}{2}, {\bf 2} \bigr]  \qquad \hbox{or} \qquad \bigl[ j=0, {\bf 1} \bigr] \oplus \bigl[ j=1, {\bf 1} \bigr]  \ , 
\ee
where $j$ denotes the spin of the $\mathfrak{sl}(2,\mathds{R})$ representation with $C^{\mathfrak{sl}(2,\mathds{R})}=-j(j-1)$. Including the action of the supercharges, we therefore conclude that the only possible highest weight representation is
\be \label{shortrep}
\begin{tabular}{ccc}
& $(\mathscr{C}^{\frac{1}{2}}_\lambda,\mathbf{2})$ & \\
$(\mathscr{C}^{1}_{\lambda+\frac{1}{2}},\mathbf{1})$ & & $(\mathscr{C}^{0}_{\lambda+\frac{1}{2}},\mathbf{1})$
\end{tabular} 
\ee
Here $\mathscr{C}^{j}_\lambda$ denotes the $\mathfrak{sl}(2, \mathbb{R})$ continuous series representation labelled by the spin $j$, as above, together with $\lambda$, the $J^3_0$ eigenvalue (modulo 1). Note that while $j$ is fixed, the $J^3_0$ eigenvalue given in \eqref{eigenvs} is not determined by the above considerations. This is fixed in the full string theory by the mass shell, i.e.\ the Virasoro conditions. 

This reproduces the result of \cite{Eberhardt:2018ouy}, see in particular eq.~(4.2). 
The fact that at $k=1$ we only have these supersymmetric short multiplets  was crucial to matching the physical spectrum of the string theory with that of the symmetric product orbifold of $\mathbb{T}^4$. In particular, it implies the absence of the continuous or ``long string" states which are a feature of the string theory (with pure NS-NS flux) for $k>1$. In contrast, here we only have the state at the bottom of the continuum $j=\tfrac{1}{2}+ip$ with $p=0$, together with its superpartners. The free field representation of the $k=1$ theory automatically captures these special properties. 

Note that the supercharge generators involve one complex fermion and one symplectic boson and change therefore both $U_0$ and $V_0$ by $\pm \frac{1}{2}$, while maintaining $Z_0=0$; they therefore map between the representation in the first line of (\ref{shortrep}), and the two representations in the second line and vice versa. 

\subsection{Spectral flow}

For the actual world-sheet theory we do not just need these highest weight representations, but also the representations that are obtained from them by spectral flow. As was already explained in \cite{Eberhardt:2018ouy}, there are two spectral flow actions one can define on these free fields, namely 
\begin{subequations}
\begin{align}
\sigma^{(+)} (\eta^+_r)&=\eta^+_{r-\frac{1}{2}}\ , & \sigma^{(-)} (\xi^+_r)&=\xi^+_{r-\frac{1}{2}}\ , \label{eq:spectral flow 1}\\
\sigma^{(+)} (\xi^-_r)&=\xi^-_{r+\frac{1}{2}}\ , & \sigma^{(-)} (\eta^-_r)&=\eta^-_{r+\frac{1}{2}}\ , \label{eq:spectral flow 2}
\end{align}
\end{subequations}
on the symplectic bosons, and 
\begin{subequations}
\begin{align}
\sigma^{(+)} (\chi^+_r)&=\chi^+_{r+\frac{1}{2}}\ , & \sigma^{(-)} (\psi^+_r)&=\psi^+_{r+\frac{1}{2}}\ , \label{eq:spectral flow 1f}\\
\sigma^{(+)} (\psi^-_r)&=\psi^-_{r-\frac{1}{2}}\ , & \sigma^{(-)} (\chi^-_r)&=\chi^-_{r-\frac{1}{2}}\ ,\label{eq:spectral flow 2f}
\end{align}
\end{subequations}
on the free fermions. The combination 
\be
\sigma = \sigma^{(+)} \circ \sigma^{(-)} 
\ee
then acts as the usual spectral flow automorphism on $\mathfrak{psu}(1,1|2)_1$, and its action on the $\mathfrak{psu}(1,1|2)$ generators is spelled out in eqs.~(\ref{eq:spectral flow a}) -- (\ref{eq:spectral flow e}). This spectral flow leaves the $\mathfrak{u}(1)$ generators 
\be
\sigma (U_m) = U_m \ , \qquad \sigma(V_m) = V_m 
\ee
invariant. The other natural combination is 
\be
\hat{\sigma} = \sigma^{(+)} \circ \bigl( \sigma^{(-)}  \bigr)^{-1} 
\ee
and it acts on the various fields as described in Appendix~\ref{app:W}, see eqs.~(\ref{Jhat}) -- (\ref{Vhat}).

Finally we need to fix our conventions for how to describe spectrally flowed representations. 
Suppose that $\tau$ is some spectral flow automorphism, i.e.\ some combination of $\sigma^{(\pm)}$, then we define the $\tau$-spectrally flowed representation to be spanned by the  states $[\Phi]^\tau$, where $\Phi$ is a state in a highest weight representation of the kind described in the previous subsection, and the action of any generator $A_n$ on $[\Phi]^\tau$ is defined by 
\be\label{specflow}
A_n \, [\Phi]^\tau  \equiv [ \tau(A_n)\, \Phi ]^\tau \ . 
\ee
Typically, the spectrally flowed representations are not highest weight representations with respect to $\mathfrak{psu}(1,1|2)$, see e.g.\ the discussion in Appendix~\ref{app:twisted}. However, this is not always the case. In particular, the $\hat{\sigma}^2$-spectrally flowed vacuum representation is again the vacuum representation with respect to $\mathfrak{psu}(1,1|2)$ since 
\be\label{Wdef}
|0\rangle^{(1)} = \Bigl[ \psi^+_{-3/2} \, \psi^-_{-3/2} \,  \psi^+_{-1/2} \,  \psi^-_{-1/2} |0\rangle \Bigr]^{\hat{\sigma}^2} \ , 
\ee
defines the vacuum state with respect to $\mathfrak{psu}(1,1|2)$. (This is explained in more detail in Appendix~\ref{app:W}.) However --- and this will be important below --- its eigenvalues with respect to $U_0$ and $V_0$ are now different since we have 
\be\label{Wcharges}
U_0 \, |0\rangle^{(1)} = |0\rangle^{(1)} \ , \qquad V_0 \, |0\rangle^{(1)} = - |0\rangle^{(1)} \ , \qquad 
Z_0 \, |0\rangle^{(1)} = 0 \ . 
\ee
In the following we shall denote the corresponding vertex operator by 
\be\label{Wfield}
W(z) \equiv V(|0\rangle^{(1)},z) \ . 
\ee

\subsection[Vertex operators in the \texorpdfstring{$x$}{x}-basis]{\boldmath Vertex operators in the \texorpdfstring{$x$}{x}-basis}

As in \cite{Eberhardt:2019ywk} we are interested in calculating the correlation functions of the world-sheet vertex operators corresponding to the twisted sector ground states. The vertex operators depend as usual on $z$, the coordinate on the world-sheet, but it is natural to introduce also a coordinate $x$ for the dual CFT (on the boundary of ${\rm AdS}_3$) and to consider the vertex operators 
\be
V^{w}(\Phi;x,z) =  e^{ z L_{-1}} e^{x J^+_0} \, V^{w}(\Phi;0,0) \, e^{-x J^+_0} \, e^{- z L_{-1}} \ ,
\ee
where $w$ denotes the spectral flow with respect to $\sigma^w$. 
Here we have used that the translation operator on the world-sheet is $L_{-1}$, while that in the dual CFT is $J^+_0$; these two operators commute with one another, and therefore the definition of these vertex operators is unambiguous. Furthermore, we define the action of $V(\Phi;0,0)$ on the vaccum of the world-sheet theory to be the state $\Phi$, 
\be
V^{w}(\Phi;0,0) \, |0\rangle = [\Phi]^{\sigma^w} \ . 
\ee
Because of locality, this then fixes the definition of the vertex operator completely, see e.g.\ \cite{Goddard:1989dp,Gaberdiel:1998fs}.

It follows from (\ref{specflow}) that the OPE of the free fields $\eta^\pm(z)$ and $\xi^\pm(z)$ near a $\sigma^w$-spectrally flowed highest weight state\footnote{These are the states that correspond to twisted sector ground states, see the analysis in Appendix~\ref{app:twisted}. Note that while the relevant states before spectral flow have fermionic descendants, they are indeed highest weight with respect to the symplectic bosons, and this is the only thing that matters in the following.} is of the form 
\begin{eqnarray}
\eta^\pm(z)\, [\Phi]^{\sigma^w} & = & \sum_{r \leq \pm \frac{w}{2}} [\eta^\pm_{r\mp\frac{w}{2}}\Phi]^{\sigma^w} \, z^{-r-\frac{1}{2}}  \\ 
\xi^\pm(z)\, [\Phi]^{\sigma^w} & = & \sum_{r \leq \pm \frac{w}{2}} [\xi^\pm_{r\mp\frac{w}{2}}\Phi]^{\sigma^w} \, z^{-r-\frac{1}{2}} \ , 
\end{eqnarray}
where $r\in \mathbb{Z} + \tfrac{1}{2}$ if $w$ is odd, and $r\in \mathbb{Z}$ for $w$ even (so that the modes that act on the states before spectral flow are always integer moded). In particular, if $\Phi$ is a highest weight state of the form $|m_1,m_2\rangle$, then $\eta^+$ and $\xi^+$ have poles of order $\frac{w+1}{2}$, 
\begin{eqnarray}
\eta^+(z) \, \bigl[|m_1,m_2\rangle\bigr]^{\sigma^w} & \sim &  z^{- \frac{w+1}{2}}  \bigl[2 m_1\, |m_1+\tfrac{1}{2},m_2\rangle\bigr]^{\sigma^w} + {\cal O}\bigl(z^{- \frac{w-1}{2}}\bigr)  \\
\xi^+(z) \, \bigl[|m_1,m_2\rangle\bigr]^{\sigma^w} & \sim &  z^{- \frac{w+1}{2}}  \bigl[|m_1,m_2+\tfrac{1}{2}\rangle\bigr]^{\sigma^w} + {\cal O}\bigl(z^{- \frac{w-1}{2}}\bigr) \ , \label{OPE2}
\end{eqnarray}
where we have used (\ref{xizero}). On the other hand, $\eta^-$ and $\xi^-$ have zeros of order $\frac{w-1}{2}$,
 \begin{eqnarray}
\eta^-(z) \, \bigl[|m_1,m_2\rangle\bigr]^{\sigma^w} & \sim &  z^{\frac{w-1}{2}}  \bigl[- 2 m_2 |m_1,m_2-\tfrac{1}{2}\rangle\bigr]^{\sigma^w} + {\cal O}\bigl(z^{ \frac{w+1}{2}}\bigr)  \\
\xi^-(z) \, \bigl[|m_1,m_2\rangle\bigr]^{\sigma^w} & \sim &  z^{\frac{w-1}{2}}  \bigl[- |m_1-\tfrac{1}{2},m_2\rangle\bigr]^{\sigma^w} + {\cal O}\bigl(z^{ \frac{w+1}{2}}\bigr) \ . \label{OPE4}
\end{eqnarray}
If the corresponding vertex operator is evaluated at $x\neq 0$, there are in addition correction terms that come from 
\begin{align}\label{eq:transprop}
S(\zeta) \, V^{w}_{m_1,m_2}(x;z) & = S(\zeta) \, \mathrm{e}^{x J_0^+} \, V_{m_1,m_2}^w(0;z)\, \mathrm{e}^{- x J_0^+} \nonumber \\
& = \mathrm{e}^{x J_0^+} \bigl[ S^{(x)}(\zeta) \, V_{m_1,m_2}^w(0;z)\bigr] \mathrm{e}^{- x J_0^+} \ , 
\end{align}
where  we have used the notation $V^{w}_{m_1,m_2}(x;z) \equiv V^w(|m_1,m_2\rangle; x,z)$, and $S(\zeta)$ is either $\eta^\alpha(\zeta)$ or $\xi^\alpha(\zeta)$. The shifted fields are  
\be
\eta^{\alpha (x)}(z) \equiv e^{-x J^+_0} \eta^\alpha(z) \, e^{x J^+_0} \ , \qquad 
\xi^{\alpha (x)}(z) \equiv e^{-x J^+_0} \xi^\alpha(z) \, e^{x J^+_0} \ , 
\ee
and we find explicitly
\be
\begin{array}{rclrcl} \label{eq:xi-shifted}
\eta^{+(x)}(z) & = & \eta^+(z) \qquad 
& \xi^{+(x)}(z) & =  & \xi^+(z)  \\ 
\eta^{-(x)}(z) & = &  \eta^-(z) - x\, \eta^+(z) \qquad 
& \xi^{-(x)}(z) & = & \xi^-(z) - x\,  \xi^+(z)  \ .
\end{array}
\ee
In particular, this therefore fixes the OPEs of the symplectic boson free fields with the vertex operators $V^w_{m_1,m_2}(x;z)$. It is then possible to perform a similar analysis as in \cite{Eberhardt:2019ywk}, except that we now derive constraint and recursion relations for the correlators involving the symplectic boson fields, rather than those involving the $\mathfrak{sl}(2,\mathds{R})$ currents; this will be done explicitly in Section~\ref{sec:examples}. As we will see, the corresponding constraints are stronger than those found in \cite{Eberhardt:2019ywk} --- this was to be expected since the $\mathfrak{sl}(2,\mathds{R})$ currents are bilinears in the symplectic bosons --- and in particular, will allow us to tie down the correlation functions essentially uniquely. 

\subsection[The OPEs with \texorpdfstring{$W(z)$}{W(z)}]{\boldmath The OPEs with \texorpdfstring{$W(z)$}{W(z)}}

As will become clear below, we will also need to know the OPEs of the symplectic boson fields with the vacuum field $W(z)$, see eq.~(\ref{Wfield}). It follows from the definition of $|0\rangle^{(1)}$ (and in particular the definition of the spectral flow $\hat{\sigma}$) that the OPEs are of the form 
\begin{align}
\label{eq:wcorr}
\eta^{\pm}(\zeta) \, W(z) & \sim \frac{1}{\zeta - z} V \Bigl(\eta^\pm _{1/2} |0\rangle^{(1)} , z\Bigr)
\ , \\
\xi^\pm(\zeta) \, W(z) & \sim {\cal O} \bigl((\zeta - z)^1  \bigr)  \ . \label{OPE6}
\end{align}

\section{Correlators in the hybrid formalism}\label{sec:picturechanging}

While everything we have said up to now is quite parallel to the $\mathfrak{sl}(2,\mathds{R})$ analysis of \cite{Eberhardt:2019ywk}, there is actually one important structural difference that will play a key role in the following. The analysis of \cite{Eberhardt:2019ywk} was effectively done in the NS-R formalism, and as a consequence the level of the bosonic (decoupled) $\mathfrak{sl}(2,\mathds{R})$ algebra was $k+2=3$. The level actually played a significant role in the analysis of the correlation functions since the simple solutions that are related to the covering map only exist provided that the identity, see eq.~(1.2) of \cite{Eberhardt:2019ywk} and eq.~(1.1) of \cite{Eberhardt:2020akk}
\be\label{constraint}
\sum_{i=1}^{n} j_i = \frac{(k+2)}{2} (n-2+2g) - (3g-3+n) 
\ee
is satisfied, which is the case for $k+2=3$ and all $j_i=\frac{1}{2}$. (Here $j_i$ is the $\mathfrak{sl}(2,\mathds{R})$ spin of the highest weight state before spectral flow, and we are considering an $n$-point function on a genus $g$ world-sheet.) In our case, the level of the bosonic $\mathfrak{sl}(2,\mathds{R})$ subalgebra of $\mathfrak{psu}(1,1|2)$ is actually $k=1$, i.e.\ in (\ref{constraint}) we need to set $k+2=1$, and thus this relation  is \emph{not} satisfied for all $j_i=\frac{1}{2}$. 

The resolution of this problem has to do with the fact that we are dealing here with a supersymmetric world-sheet theory, and we need to take picture changing into account. In this technical section we describe how everything works out as expected after taking this on board, as well as the $U_0$ charge conservation. In the rest of the paper, we will only be using the final result in \eqref{correlators1} for the physical correlators; the reader may therefore skip the rest of this section on a first reading.   

\subsection{Picture changing}

In the context of the hybrid string, the world-sheet amplitudes (for $g \geq 1$), which we need to calculate are, see  eq.~(2.7) of \cite{Berkovits:1999im}
\be\label{6.1}
\int_{{\cal M}_{g,n}} \Bigl\langle \hat{G}^-(\mu_1) \cdots \hat{G}^-(\mu_{3g-3}) \Bigl[ \int \tilde{G}^+ \Bigr]^{g-1} \, \int J \, 
\prod_{i=1}^{n} \hat{G}^-_{-1} \Phi_{i} \Bigr\rangle \ , 
\ee
where we have only written out the left-moving component of the correlator, and we have already anticipated that not all ${G}^-$ fields and  $G^-_{-1} \Phi_{i}$ descendants will in fact involve $G^-_{-1}$ --- as we shall see momentarily, some of them will in fact involve $\tilde{G}^-_{-1}$, see also the comment below eq.~(2.9) and above eq.~(3.5) of \cite{Berkovits:1999im}. 

The different supercurrents can be written out explicitly in terms of the bosonised ghost variables $(\rho, \sigma)$ as well as the ${\cal N}=2$ $\mathfrak{u(}1)$ current $J=\partial H$, see the final equation of \cite{Berkovits:1999im}
\begin{align}
G^- & = e^{-i \sigma} + G_C^- \\
\tilde{G}^- & = e^{-2\rho - i \sigma - i H} {\cal Q} + e^{-\rho - i H} {\cal T} + e^{-\rho - i \sigma} \tilde{G}_C^- \\
\tilde{G}^+ & = e^{\rho+i H}+ e^{\rho+i \sigma}\tilde{G}_C^+ \ .
\end{align}
Here we have used that for the free field realisation at level $k=1$, the $(p)^4$ term vanishes, and the ${\cal Q}$ operator takes the form, see eqs.~(2.26) and (2.27) of \cite{Gerigk:2012cq}
\be
{\cal Q} = 2 ({\chi}^+ {\chi}^-) \bigl(\xi^+ \partial \xi^- - \xi^- \partial \xi^+ \bigr) \ . 
\ee
(For the physical states we are interested in, the ${\cal T}$ and the $\tilde{G}_C^-$ operators do not play a role.) On the other hand, the physical states are of the form \cite{Gerigk:2012cq}, 
\be\label{physical}
\Phi\, e^{2\rho + i \sigma + i H} \ , 
\ee
where for example for the twisted sector ground states $\Phi = [\Phi_w]^{\sigma^w}$ as in Appendix~\ref{app:twisted}. This is now sufficient to determine how many $\hat{G}^-$ in (\ref{6.1}) are $G^-$, and how many are $\tilde{G}^-$. We claim that (for $g\geq 1$) of the $n+3g-3$ $\hat{G}^-$ fields that appear altogether we take\footnote{For $g=0$ we can make sense of the negative number of $G^{-}$ fields in  (\ref{hatG}), as well as the negative power of $\tilde{G}^+$ in (\ref{6.1}), as in \cite{Berkovits:1994vy}, see the discussion around eq.~(2.23) there.}
\be\label{hatG}
(n-2+2g) \, \times \, \tilde{G}^{-}_{-1}  \ , \qquad \hbox{and} \qquad (g - 1) \, \times \, G^-_{-1} \ . 
\ee
This prescription is simply a consequence of overall charge conservation. Indeed, the total exponential of the ghost fields equals
\begin{align}
& n (2\rho + i \sigma + iH) + (n-2+2g) (-2 \rho - i \sigma - iH) +(g-1) (-i\sigma) + (g-1) (\rho+iH) \nonumber \\
& =  (g-1) \bigl( 2 (-2 \rho - i \sigma - iH)  - i\sigma + \rho + iH\bigr) \nonumber \\
& = (g-1) ( -3\rho -3 i \sigma - iH) \ ,
\end{align} 
where the last term in the first line comes from the $(g-1)$ $\tilde{G}^+$ fields. The last line describes exactly the required background charge for genus $g$. 

This prescription now reproduces correctly (\ref{constraint}): if we apply $\tilde{G}^{-}_{-1}$ instead of $G^-_{-1}$ to the physical vertex operator, we pick up the term
\be
{\cal Q}_{-1}\,  [\Phi_w]^{(w)} \ . 
\ee
(The prefactor coming from 
\be
e^{-2\rho - i\sigma -iH}(z) \, e^{2\rho + i\sigma + iH}(w) \sim (z-w)^{4-1-2} = (z-w)^1 
\ee
guarantees that we pick up the ($-1$)-mode of ${\cal Q}$.) From our analysis in Appendix~\ref{app:twisted} it is easy to see that the $-1$ mode is simply 
\be\label{Qm1odd}
{\cal Q}_{-1}\, [\Phi_w]^{(w)}  = 2 w \Bigl[ \xi^+_0 \xi^-_0 \, {\chi}^+_{\frac{w-1}{2}} \, {\chi}^-_{- \frac{w+1}{2}} \Phi_w\Bigr]^{\sigma^w} \ , 
\ee
for $w$ odd, see in particular eq.~(\ref{twistedodd}), and similarly for $w$ even, i.e.\ 
\be
{\cal Q}_{-1} [\Phi_w^+]^{(w)}  = 2 w \Bigl[ \xi^+_0 \xi^-_0 \, {\chi}^+_{\frac{w}{2}-1} \, {\chi}^-_{- \frac{w}{2}} \Phi_w^+\Bigr]^{\sigma^w} \ , 
\ee
and 
\be
{\cal Q}_{-1} [\Phi_w^-]^{(w)}  = 2 w \Bigl[ \xi^+_0 \xi^-_0 \, {\chi}^+_{\frac{w}{2}} \, {\chi}^-_{- \frac{w}{2}-1} \Phi_w^-\Bigr]^{\sigma^w} \ ,
\ee
see eqs.~(\ref{twistedevenp}) and (\ref{twistedevenm}). What is important is that in each case we have the product $\xi_0^+ \xi_0^-$ which, according to eq.~(\ref{xizero}), maps 
\be
m_1 \mapsto m_1 - \frac{1}{2} \ , \qquad m_2 \mapsto m_2 + \frac{1}{2} \ . 
\ee
As a consequence, see eq.~(\ref{J30}), the $J^3_0$ eigenvalue of this state is unchanged, but the spin is shifted by 
\be
j = m_1 - m_2 \mapsto j -  1 \ . 
\ee
 On the other hand, if we replace one of the $\hat{G}^-$ in the Beltrami differential by $\tilde{G}^-$, we need to integrate the vertex operator associated to the state $\mathcal{Q} = \mathcal{Q}_{-3} |0\rangle$. We can then express the $\xi^\pm$ modes of $\mathcal{Q}$ in terms of a contour integral, and use the techniques of Section~\ref{sec:examples} to rewrite the $\xi^\pm$ fields in terms of $\xi^\pm_0$ modes acting on the physical vertex operators. This therefore has the same effect as above, i.e.\ each $\tilde{G}^-$ field reduces the (total) spin by $1$. As a consequence, the sum over the spins $j_i$ on the left-hand-side of (\ref{constraint}) becomes (initially each spin was $j_i=\tfrac{1}{2}$) 
\be
\sum_{i=1}^{n} j_i = \frac{n}{2} - (n-2 + 2g) = \frac{1}{2} (n-2+2g) - (3g-3+n) \ , 
\ee
i.e.\ it satisfies the constraint (\ref{constraint}) with the bosonic level equal to $k+2=1$.

\subsection[\texorpdfstring{$U_0$}{U0} charge constraint]{\boldmath \texorpdfstring{$U_0$}{U0} charge constraint}\label{sec:U0charge}

While the previous consideration takes into account the constraints from picture changing, it introduces another problem: in our free field realisation, all vertex operators initially have $U_0=0$, and this is also not affected by spectral flow. However, once we have applied ${\cal Q}$, we do not only shift $j_i\mapsto j_i-1$, but also $U_0\mapsto U_0 -1$, see eq.~(\ref{U0}). As a consequence the resulting correlators vanish on the nose since they do not respect overall $U_0$ charge neutrality. (This is also something we observed in the explicit analysis of Section~\ref{sec:examples}.) 

This problem is a consequence of our specific free field realisation, and it already implicitly reared its head in the analysis of Appendix~C (and in particular eq.~(C.24)) of \cite{Eberhardt:2018ouy}. Indeed, the full spectrum of our $\mathfrak{psu}(1,1|2)$ model appears not just once in the free field realisation, but there is a copy of it for each eigenvalue of $U_0$. In calculating correlation functions we therefore need to choose the states such that the overall $U_0$ charge vanishes, and this can always be achieved. In our context, the simplest way of arranging for this is to add $(n-2+2g)$ vacuum fields with $U_0=+1$. The relevant field was already constructed above, see eq.~(\ref{Wdef}), and was denoted by $W(z) \equiv V(|0\rangle^{(1)},z)$.\footnote{Since $|0\rangle^{(1)}$ behaves as the vacuum state with respect to $\mathfrak{psu}(1,1|2)$, the corresponding vertex operator does not depend on $x$.}  Thus the correlation functions we shall be studying in the following are of the form 
\begin{equation}\label{correlators}
\Bigl\langle \,
\prod_{a=1}^L \mathcal{Q}(y_{a})
\prod_{\alpha=1}^{n-2+2g} W(u_{\alpha})\prod_{i=1}^{n}V_{m_1^i,m_2^i}^{w_i}(x_i;z_i) \, \Bigr\rangle \ , 
\end{equation}
where $(n+2g -2 - L)$ $j_j$ are  $j_j=-\tfrac{1}{2}$, while the remaining $j_i$ are equal to $j_i=\tfrac{1}{2}$. In this paper we shall concentrate on the correlators on the sphere, $g=0$; then we may take $L=0$, and hence consider the correlation functions 
\begin{equation}\label{correlators1}
\Bigl\langle \,
\prod_{\alpha=1}^{n-2} W(u_{\alpha})\prod_{i=1}^{n}V_{m_1^i,m_2^i}^{w_i}(x_i;z_i) \, \Bigr\rangle \ , 
\end{equation}
where $(n-2)$ $j_j$ are  $j_j=m_1^j - m_2^j = -\tfrac{1}{2}$, while the remaining $j_i$ equal $j_i=m_1^i - m_2^i = \tfrac{1}{2}$.

\section{A covering map identity and its consequences}\label{sec:twistor}

In this section we begin with analysing the correlation functions of the form (\ref{correlators1}) for the case of the sphere ($g=0$). One way to constrain these correlators is to repeat the analysis of \cite{Eberhardt:2019ywk}, where instead of determining constraint and recursion relations for the correlators involving the $\mathfrak{sl}(2,\mathds{R})$ currents, we now replace these currents by the symplectic bosons. While this is a possible avenue --- and we shall spell it out in some detail in Section~\ref{sec:examples} --- there is actually a much more elegant method that gives rise to essentially the same constraints. Apart from being a computationally powerful idea, it also sheds light on the conceptual underpinning of our analysis; in particular, it suggests that we should think of our free field realisation as giving rise to a twistor-like string theory.

\subsection{The covering map}\label{sec:covering}

Before we begin with the detailed analysis of our central relation, let us remind the reader about some properties of covering maps. We will restrict ourselves here to covering maps of Riemann surfaces with $g=0$.

Let $\Gamma:\text{S}^2\to\text{S}^2$ be a holomorphic function from the Riemann sphere to itself. Given a collection of points $\{z_i\}$ and $\{x_i\}$ on $\text{S}^2$ and a set of positive integers $w_i$ for $i=1,\ldots,n$, we say that $\Gamma$ is a \textit{branched covering map} with branching indices $\{w_i\}$ if
\begin{equation}
\Gamma(z)\sim x_i+\mathcal{O}\bigl((z-z_i)^{w_i}\bigr)\ ,\qquad z\to z_i 
\end{equation}
for all $i$, with no other critical points, i.e.\ the only zeroes of $\partial\Gamma(z)$ are at $z=z_i$. A fundamental result in the theory of Riemann surfaces is the Riemann-Hurwitz formula, which, for genus $g=0$ surfaces, states that the order of a branched covering map (i.e., the number of preimages of a generic point) can be determined purely by the branching indices, and is given explicitly as
\begin{equation}
N=1+\sum_{i=1}^{n}\frac{w_i-1}{2}\ .
\label{eq:Riemann-Hurwitz}
\end{equation}
In what follows we will be mostly concerned with the conditions for the existence and uniqueness of a covering map. A necessary condition for existence is that the $w_i$ satisfy the selection rules
\begin{equation}
\sum_{i\neq j}^{n}(w_i-1)\geq w_j-1\ ,\qquad \sum_{i=1}^{n}(w_i-1)\in 2\mathbb{Z}\ ,
\label{eq:selection-rules}
\end{equation}
where the first statement holds for all $j$, and the second statement is merely the requirement that the order \eqref{eq:Riemann-Hurwitz} of $\Gamma$ is an integer. These conditions are enough to guarantee existence for $n=3$, but for $n\geq 4$ a covering map generically does not exist. To illustrate this, note that $\Gamma$, as a holomorphic function from the sphere to itself, can be expressed as a ratio of polynomials of order $N$ (assuming that $\infty$ is not a branching point). That is, we can define polynomials $p^{\pm}(z)$ such that
\begin{equation}
\Gamma(z)=-\frac{p^-(z)}{p^+(z)}\ .
\label{eq:gamma-def}
\end{equation}
The requirement, then, that $\Gamma(z)$ has a critical point of order $w_i$ at $z=z_i$, whose image is $x_i$, can be written as
\begin{equation}
p^-(z)+x_i\,p^+(z) = \mathcal{O}\bigl((z-z_i)^{w_i}\bigr)\ .
\label{eq:gamma-constraint}
\end{equation}
This can be thought of as a linear homogeneous system in the $2N+2$ coefficients of the polynomials $p^{\pm}(z)$. There are $\sum_{i}w_i=2N+n-2$ such equations, and thus a solution will only exist on some $n-3$ dimensional subspace of the configuration space labeled by $\{z_i\},\{x_i\}$. (Note that the overall scale of the polynomials $p^{\pm}(z)$ drops out for the determination of $\Gamma(z)$, and thus there are only $2N+1$ meaningful coefficients.)

It will be useful, for what follows in the next subsection, to view these conditions in a slightly different  way. First, note that \eqref{eq:gamma-constraint} as well as its derivative with respect to $z$ can be used to eliminate the $x_i$. Next we observe that the `Wronskian' behaves in the vicinity of the critical point $z_i$ as, see also \cite{Lunin:2001,Pakman:2009,Roumpedakis:2018tdb} 
\be
p^{-\prime}(z)p^+(z)-p^{+\prime}(z)p^-(z)  =  \mathcal{O}\bigl((z-z_i)^{w_i-1}\bigr) \ ,
\ee
where the prime denotes the derivative. (This identity can also be understood as arising from the numerator of the rational function $\partial\Gamma(z)$.) 
Since the left hand side is, in fact, a polynomial of degree $2N-2$,\footnote{Naively it might appear to be one of degree $2N-1$, however the leading power has an identically vanishing coefficient.} we conclude that 
\begin{equation}
p^{-\prime}(z)p^+(z)-p^{+\prime}(z)p^-(z) = C\prod_{i=1}^n(z-z_i)^{w_i-1}\ .
\label{eq:ppm-constraint}
\end{equation}
Here we have used that the right hand side of \eqref{eq:ppm-constraint} has the correct vanishing behaviour at each of the critical points, and that it is also of degree $2N-2$ because of  \eqref{eq:Riemann-Hurwitz}; it must therefore agree with the left hand side up to an overall constant which we have denoted by $C$. The important point to note is that all the $x_i$ have dropped out of \eqref{eq:ppm-constraint}. 

A short argument (see Appendix \ref{app:cover}) then shows that \eqref{eq:ppm-constraint} determines the polynomials $p^{\pm}(z)$ in terms of any two of its coefficients, as well as one overall scale factor for each polynomial, see eq.~\eqref{eq:ppm-sol}.
The overall scale common to both $p^{\pm}(z)$ drops out as discussed above. So we are left with one relative scale factor and two ratios of coefficients as unknowns. These three can be fixed by going back to \eqref{eq:gamma-constraint} for three of the branch points, i.e.\ we may use  
\begin{equation}
p^-(z_i)+x_i\,p^+(z_i) = 0
\label{eq:gamma-constr-sp0}
\end{equation}  
for, say, $i=1,2,3$. This fixes the $p^{\pm}(z)$ completely, up to an overall common factor (and up to discrete choices). 

In terms of the branched cover $\Gamma(z)$ defined in \eqref{eq:gamma-def}, we can now rewrite the remaining equations of \eqref{eq:gamma-constraint} (i.e.\ for $i=4\ldots n$) as
\begin{equation}
\bigl(x_i-\Gamma(z)\bigr)\,p^+(z_i) = \mathcal{O}\bigl((z-z_i)^{w_i}\bigr)\ .
\label{eq:gamma-rest}
\end{equation}
In particular, if $x_i\neq \Gamma(z_i)$, then the covering map does not exist as specified (i.e.\ with $p^+(z_i)\neq 0$). Thus there are $n-3$ conditions ($x_i = \Gamma(z_i)$ for $i=4,\ldots,n$) that need to be satisfied in order for the covering map to exist.

\subsection{The incidence relation and delta function localisation}\label{sec:twistorrel}

With these preparations at hand we can now turn to the main claim of this paper, which constrains the structure of the correlation functions in the symplectic boson theory almost completely. Let $\{z_i\}$ and $\{x_i\}$ be some marked points on the world-sheet and boundary ${\rm S}^2$ respectively, and let $\{w_i\}$ be a set of branching indices. Assuming that a covering map $\Gamma(z)$ for this configuration exists, we claim that
\begin{equation}
\Bigl\langle  \Bigl(\xi^-(z)+\Gamma(z)\xi^+(z)\Bigr)\, 
\prod_{\alpha=1}^{n-2}W(u_{\alpha})\prod_{i=1}^{n}V_{m_1^i,m_2^i}^{w_i}(x_i;z_i) \Bigr\rangle =0\ .
\label{eq:magic}
\end{equation}
Furthermore, we claim that {\em if such a covering map does not exist}, then the correlation functions vanish identically, 
\begin{equation}
\Bigl\langle  \prod_{\alpha=1}^{n-2}W(u_{\alpha})\prod_{i=1}^{n}V_{m_1^i,m_2^i}^{w_i}(x_i;z_i) \Bigr\rangle =0 \ .
\label{eq:vanishing}
\end{equation}
In fact, as we will see below, the correlators \eqref{eq:vanishing} are delta function localised on the locus of the covering map, see eq.~(\ref{corrstructfinal}).
\smallskip

In order to prove \eqref{eq:magic}, we begin by defining the functions
\begin{equation}
P^{\pm}(z)=\frac{\prod_{i=1}^{n}(z-z_i)^{\frac{w_i+1}{2}}}{\prod_{\alpha=1}^{n-2}(z-u_{\alpha})}\Bigl\langle \xi^\pm(z)\prod_{\alpha=1}^{n-2}W(u_{\alpha})\prod_{i=1}^{n}V_{m_1^i,m_2^i}^{w_i}(x_i;z_i) \Bigr\rangle \ .
\label{polynomial-def}
\end{equation}
The prefactors have been chosen so as to ensure that these functions are globally well-defined, i.e.\ do not have any branch cuts. (Since the states $\ket{m_1,m_2}$ define a Ramond-sector representation with respect to the symplectic bosons, the symplectic bosons pick up a sign in going around the corresponding states for even $w$.) Furthermore, because of \eqref{OPE2} and $\eqref{OPE4}$, as well as \eqref{OPE6}, the functions $P^{\pm}(z)$ have no poles at $z=z_i$ or $z=u_{\alpha}$ (and thus no poles for any finite $z$). Finally, because of the prefactor and the $\frac{1}{z}$ fall-off behavior of $\xi^{\pm}(z)$, the functions $P^{\pm}(z)$ have the asymptotic form
\begin{equation}
P^{\pm}(z)\sim z^N,\hspace{0.5cm}z\to\infty\ ,\qquad \text{where}\quad 
N=1+\sum_{i=1}^{n}\frac{w_i-1}{2}\ .
\end{equation}
It would thus seem that the $P^{\pm}(z)$ are polynomials in $z$ of order $N$, but there is one subtlety that we need to discuss: the above argument only describes the dependence on $z$, but the coefficients $z^k$ of $P^{\pm}(z)$ could be arbitrary distributions of the other variables $z_i$ and $x_i$. (In fact, this is not just an academic possibility, but as we will see, they actually turn out to be such distributions.) However, the distributional nature of $P^{\pm}(z)$ is of a special kind: the coefficients of both polynomials $P^{\pm}(z)$ are all proportional (with proportionality factors that are all regular functions of $z_i$ and $x_i$) to a single coefficient, say that of $z^0$  in $P^+(z)$, which by itself can be (and will be) a distribution $D(z_i, x_i)$ of $z_i$ and $x_i$.\footnote{The overall distribution could, in principle, also depend on the $u_{\alpha}$ but the dependence of the correlators on $u_\alpha$ is completely fixed by conformal symmetry, as the $W$-fields do not carry $\mathfrak{psu}(1,1|2)$ charge, see the discussion below eq.~\eqref{ufactor}.} Thus we can write
\begin{equation}
P^{\pm}(z) = D(z_i, x_i)\, \hat{p}^{\pm}(z) \ , 
\label{eq:Pdistr}
\end{equation}
where $\hat{p}^{\pm}(z)$ are now polynomials in $z$ of degree $N$ whose coefficients are ordinary functions of $(z_i, x_i)$. 

In order to show (\ref{eq:Pdistr}) we note that the OPEs of $\xi^\pm(z)$ with the fields $V^{w_i}_{m_1^i,m_2^i}(x_i;z_i)$, see eqs.~\eqref{OPE2}, \eqref{OPE4}, and \eqref{eq:xi-shifted}, imply  that the $P^{\pm}(z)$ satisfy the constraints
\begin{equation}
P^-(z)+x_i \, P^+(z)\sim \mathcal{O} \bigl((z-z_i)^{w_i} \bigr)\ ,\qquad z\to z_i \ . 
\label{polynomial-constraint2}
\end{equation}
Note that these are essentially the same equations that characterise the covering map, see eq.~(\ref{eq:gamma-constraint}). In particular, they define $2N+n-2$ homogeneous linear equations (with coefficients that are regular functions of the $(z_i,x_i)$) for the $2N+2$ coefficients of the polynomials $P^\pm(z)$. Given the structure of the covering maps we know that we can choose a (subset) of $2N+1$ equations such that the resulting system becomes non-degenerate; thus, all coefficients of $P^\pm(z)$ are proportional (with proportionality constants that are regular functions of $(z_i,x_i)$) to one coefficient. This establishes eq.~(\ref{eq:Pdistr}).

Next writing $P^\pm(z)$ in terms of eq.~\eqref{eq:Pdistr} we note that the polynomials $\hat{p}^{\pm}(z)$ obey
\be
D(z_i, x_i)\, \Bigl( \hat{p}^-(z)+x_i\,\hat{p}^+(z) \Bigr)= \mathcal{O}\bigl((z-z_i)^{w_i}\bigr)\ .
\label{eq:branching}
\ee
On the support of the distribution $D(z_i, x_i)$ it then follows from the arguments below \eqref{eq:ppm-constraint} and in Appendix~\ref{app:cover}, that we can determine $\hat{p}^{\pm}(z)$ up to two unknowns and the two overall scale factors. By imposing the conditions 
\begin{equation}
\hat{p}^-(z_i)+x_i \, \hat{p}^+(z_i) =0  
\label{polynomial-constraint}
\end{equation}
for three of the $i$, say $i=1,2,3$, all the unknowns except for the overall common scale factor of $\hat{p}^{\pm}(z)$ can be determined. This overall factor, which is not fixed by \eqref{polynomial-constraint}, can be absorbed into the distribution $D(x_i, z_i)$. Using again eq.~\eqref{eq:Pdistr} we therefore find 
\begin{equation}
P^-(z)+\Gamma(z)P^+(z)=0\ ,
\label{magic-polynomials}
\end{equation}
where $\Gamma(z)$ is the `branched cover' defined in \eqref{eq:gamma-def}, i.e.\ it is the function that has  branching index $w_i$ at $z_i$ and satisfies $\Gamma(z_i) = x_i$ for $i=1,2,3$, 
and we have used that $p^\pm(z)= \tilde{c} \, \hat{p}^\pm(z)$ for some constant $\tilde{c}$. If $\Gamma(z)$ is in fact the actual branched cover, i.e.\ if also $\Gamma(z_i) = x_i$  for $i=4,\ldots,n$, then eq.~(\ref{magic-polynomials}) is equivalent to the incidence relation of eq.~\eqref{eq:magic}, where we have used the definitions in \eqref{polynomial-def}. This proves our first claim. 

It remains to show that the correlators \eqref{eq:vanishing} vanish if the covering map does not exist. We will actually show the stronger statement that the correlators are delta function localised on the locus of the covering map. To see this we evaluate \eqref{magic-polynomials} at $z=z_i$ and use (\ref{eq:Pdistr}) and (\ref{eq:branching}) to conclude that  
\begin{equation}
D(x_i,z_i)\, \Bigl( (x_i -\Gamma(z_i))\, \hat{p}^+(z_i) \Bigr)  =0  \qquad  (i=1,\ldots, n) \ .
\label{delta-constraint}
\end{equation}
By construction of the $\hat{p}^\pm(z)$, this identity is identically true for $i=1,2,3$, but for $i\geq 4$ it leads to the condition 
\be\label{deltacon}
D(x_i,z_i)\, \Bigl( x_i -\Gamma(z_i) \Bigr)  =0 \qquad (i=4,\ldots,n) \ ,
\ee
where we have used that $\hat{p}^+(z_i)\neq 0$. Thus it follows that\footnote{We are assuming here that the distribution $D(x_i,z_i)$ can be expressed in terms of delta functions and their derivatives. Then the statement follows from the fact that, within this particular space of distributions, the unique solution to the equation $x f(x)=0$ is $f(x)\propto\delta(x)$.}
\be
D(x_i,z_i) = \sum_\Gamma \hat{C}_\Gamma \, \prod_{i=4}^{n} \, \delta\bigl( x_i -\Gamma(z_i) \bigr) \ , 
\ee
where the sum runs over the different `covering maps', see the discussion below eq.~(\ref{magic-polynomials}), and the $\hat{C}_\Gamma$ are ordinary functions of their arguments.

Both $P^\pm(z)$ are proportional to this distribution,  because of eq.~(\ref{eq:Pdistr}). On the other hand, $P^+(z_i)$ is essentially a generic correlator of the kind we are interested in since
\begin{equation}
\lim_{z\to z_i}\frac{\prod_{\alpha=1}^{n-2}(z-u_{\alpha})}{\prod_{j\neq i}^{n}(z-z_j)^{\frac{w_j+1}{2}}}P^+(z)= \Bigl\langle \prod_{\alpha=1}^{n-2}W(u_{\alpha})\, V_{m_1^i,m_2^i+\frac{1}{2}}^{w_i}(x_i;z_i)\, \prod_{j\neq i}^{n}V_{m_1^j,m_2^j}^{w_j}(x_j;z_j) \Bigr\rangle \ .
\label{polynomial-to-correlator}
\end{equation}
Since the values of the labels $m_1,m_2$ are arbitrary,\footnote{They only have to satisfy the constraint that comes from the $U_0$ charge neutrality condition, but since $\xi^+$ also carries $U_0$ charge, this is precisely what is responsible for this shift of $m_2^i$.} it follows that the correlator itself is delta-function localised, i.e.\ that 
\begin{equation}\label{corrstructfinal}
\Bigl\langle \prod_{\alpha=1}^{n-2}W(u_{\alpha})\prod_{i=1}^{n}V_{m_1^i,m_2^i}^{w_i}(x_i;z_i) \Bigr\rangle=\sum_{\Gamma}C_{\Gamma}\,\prod_{i=4}^n \delta\left(x_i-\Gamma(z_i)\right)\ ,
\end{equation}
where $C_{\Gamma}$ contains the dependence on the $z_i$ and the $u_\alpha$ (though the dependence on the latter is rather trivial, see eq.~(\ref{ufactor}) below). We also recall that $\Gamma(z)$ here is the function that has branching index $w_i$ at $z_i$ and satisfies $\Gamma(z_i)=x_i$ for $i=1,2,3$. 

Eq.~(\ref{corrstructfinal}) therefore provides an explicit and manifest realisation of the localisation idea conjectured in \cite{Pakman:2009}, for which strong evidence was given in \cite{Eberhardt:2019ywk}. We have also worked this out explicitly in some simple examples, see Appendix~\ref{app:explicit} for more details. Finally, for four-point functions one can give a fairly direct general argument for the structure (\ref{corrstructfinal}); this is explained in Appendix~\ref{app:4ptdelta}.

\subsection{Fusion rules}\label{sec:fusion}

The identity and the proof we provided above have some immediate implications for the structure of the correlation functions. First, let $n=3$. Then \eqref{eq:vanishing} tells us that the three-point function
\begin{equation}
\Bigl\langle W(u)\prod_{i=1}^{3}V^{w_i}_{m_1^i,m_2^i}(x_i;z_i)\Bigr\rangle  
\end{equation}
is nonzero only if, see eq.~(\ref{eq:selection-rules})
\begin{equation}
w_i+w_j\geq w_k+1\ ,\qquad w_1+w_2+w_3\in 2\mathbb{Z}+1\ ,
\label{eq:fusion-rules}
\end{equation}
where the first statement holds for any distinct choices of $i,j,k$. Since this is a necessary condition for the three-point function to be non-vanishing, we conclude that \eqref{eq:fusion-rules} are (part of the) fusion rules of the theory. In fact, this reproduces precisely the expected fusion rules of the dual symmetric orbifold theory \cite{Eberhardt:2018ouy}.

\subsection{Constraining the correlation functions}

The relation \eqref{eq:magic} can also be used to constrain the $m_j^i$-dependence of the correlation functions. Assuming a covering map exists, let $a_i^\Gamma$ be the constants (that still depend on the $z_i$ and $x_i$) such that
\begin{equation}
\Gamma(z)\sim x_i+a_i^{\Gamma}(z-z_i)^{w_i}+\cdots\ ,\qquad z\to z_i\ .
\end{equation}
Now, let us consider the Laurent expansion of \eqref{eq:magic} near $z=z_i$. Using the OPEs \eqref{OPE2}, $\eqref{OPE6}$, as well as $\eqref{eq:xi-shifted}$, we have
\begin{equation}
\begin{split}
&\Bigl\langle \left(\xi^-(z)+\Gamma(z)\, \xi^+(z)\right)\prod_{\alpha=1}^{n-2}W(u_{\alpha})\prod_{i=1}^{n}V_{m_1^i,m_2^i}^{w_i}(x_i;z_i) \Bigr\rangle\\
&\hspace{2cm}\sim -(z-z_i)^{\frac{w_i-1}{2}}\Bigl\langle \prod_{\alpha=1}^{n-2}W(u_{\alpha})\, V_{m_1^i-\frac{1}{2},m_2^i}^{w_i}(x_i;z_i)\, \prod_{i\neq j}^{n}V_{m_1^i,m_2^i}(x_i;z_i) \Bigr\rangle\\
&\hspace{2.5cm}+a_i^\Gamma\left(z-z_i\right)^{\frac{w_i-1}{2}}\Bigl\langle \prod_{\alpha=1}^{n-2}W(u_{\alpha})\, V_{m_1^i,m_2^i+\frac{1}{2}}^{w_i}(x_i;z_i)\, \prod_{i\neq j}^{n}V_{m_1^i,m_2^i}(x_i;z_i) \Bigr\rangle+\cdots\ .
\end{split}
\end{equation}
Since this expression must vanish identically by \eqref{eq:magic}, we have the recursion relation
\begin{equation}
\begin{split}
&\Bigl\langle \prod_{\alpha=1}^{n-2}W(u_{\alpha})\, V_{m_1^i-\frac{1}{2},m_2^i}^{w_i}(x_i;z_i)\, \prod_{i\neq j}^{n}V_{m_1^i,m_2^i}(x_i;z_i) \Bigr\rangle \\
&\hspace{2cm}=a_i^\Gamma\, \Bigl\langle \prod_{\alpha=1}^{n-2}W(u_{\alpha})\, V_{m_1^i,m_2^i+\frac{1}{2}}^{w_i}(x_i;z_i)\, \prod_{i\neq j}^{n}V_{m_1^i,m_2^i}(x_i;z_i) \Bigr\rangle\ .
\end{split}
\label{magic-recursion}
\end{equation}
Recalling that $h_i=m_1^i+m_2^i + \frac{w_i}{2}$ is the $J^3_0$ eigenvalue (and therefore the spacetime conformal dimension of the corresponding operator in the spacetime CFT) of $\ket{m_1^i,m_2^i}$, see eqs.~(\ref{J30}) and (\ref{eq:spectral flow a}), the solution to this recursion relation is simply given by
\begin{equation}
\Bigl\langle \prod_{\alpha=1}^{n-2}W(u_{\alpha})\prod_{i=1}^{n}V_{h_i}^{w_i}(x_i;z_i) \Bigr\rangle \propto C(j_i)\prod_{i=1}^{n}\left(a_i^\Gamma\right)^{-h_i}\ ,
\end{equation}
where $C(j_i)$ is a function of the $\mathfrak{sl}(2,\mathbb{R})$ spins $j_i=m_1^i-m_2^i$, and we have written 
\be
V_{h_i}^{w_i}(x_i;z_i)  \equiv V_{m_1^i,m_2^i}^{w_i}(x_i;z_i) 
\ee
with $h_i=m_1^i+m_2^i + \frac{w_i}{2}$. This is exactly the structure of the correlation function of twisted sector ground states in the symmetric product orbifold found in \cite{Lunin:2001,Pakman:2009,Eberhardt:2019ywk, Dei:2020}. 
Putting everything together, we therefore conclude that the correlators are of the form 
\begin{equation}
\Bigl\langle \prod_{\alpha=1}^{n-2}W(u_{\alpha})\prod_{i=1}^{n}V_{h_i}^{w_i}(x_i;z_i) \Bigr\rangle =\sum_{\Gamma}W_{\Gamma}(z_i, u_{\alpha},j_i)\prod_{i=1}^{n}\left(a_i^{\Gamma}\right)^{-h_i}\, \prod_{i=4}^{n}\delta(x_i-\Gamma(z_i))\ ,
\label{general-correlator}
\end{equation}
where $W_{\Gamma}$ is some unknown function of the insertion points and $\mathfrak{sl}(2,\mathbb{R})$ spins. The correlator thus has support only on the locus where the covering map $\Gamma$ exists. Finally, we sum over all possible covering maps for a given set of branching indices $w_i$. This is exactly the analogue, in the hybrid formalism, of eq.~(5.28) of \cite{Eberhardt:2019ywk}, which was obtained there by assuming a particular solution of the $\mathfrak{sl}(2,\mathbb{R})_k$ Ward identities. The above derivation using the free field realisation has allowed us to dispense with the assumption that the delta function localised solution of the Ward identities is indeed the correct solution: at $k=1$ it is the only solution.

In  \cite{Eberhardt:2019ywk}, the delta function constraint was combined with a semi-classically exact saddle point of the classical sigma model, which gave a weight of $e^{-S_{\text{L}}[\phi_{cl}]}$ for each such covering map. Here 
\be\label{liouv}
S_\text{L}[\phi] = \frac{c}{48\pi} \int \mathrm{d}^2 z\,  \sqrt{g}\, \Bigl(2 \, \partial \phi \, \bar\partial \phi+R\, \phi\Bigr) 
\ee
is the classical Liouville action and the scale factor $\phi_{cl}$ is the conformal factor arising from the covering map, i.e. 
\begin{equation}
\phi_{cl}(z) = \ln{(|\partial{\Gamma(z)}|^2)} \ . 
\end{equation}
The action $S_\text{L}[\phi_{cl}]$ is to be evaluated in a suitably regularised way \cite{Lunin:2001}. 
Together with the results of \cite{Dei:2020}, this then suggests that the full correlator (combining both chiral and anti-chiral degrees of freedom) is  of the form
\begin{equation}\label{full-corr}
\Bigl\langle \prod_{\alpha=1}^{n-2}W(u_{\alpha})\prod_{i=1}^{n}V_{h_i}^{w_i}(x_i;z_i) \Bigr\rangle =\sum_{\Gamma}\tilde{W}_{\Gamma}\, e^{-S_L[\phi_{cl}]}\prod_{i=1}^{n}|a_i^{\Gamma}|^{-2(h_i-h_i^0)}\, \prod_{i=4}^{n}\delta^{(2)}(z_i-\Gamma^{-1}(x_i)) \ ,
\end{equation}
where we have, as always, assumed that $x_i = \Gamma(z_i)$ for $i=1,2,3$ (and $\Gamma(z)$ has branching index $w_i$ at $z_i$). Here $h^0=\frac{(w^2-1)}{4w}$ is the spacetime conformal dimension of the $w$-th spectrally flowed ground state --- the contribution from the ground state is already captured by the exponential of the classical Liouville action.\footnote{In the discussion of this paper we have assumed that the states are spectrally flowed highest weight states with respect to the $\mathfrak{sl}(2,\mathds{R})$ factor, but their excitation with respect to the other degrees of freedom can be arbitrary. One would expect that one can bring all physical states into this `gauge', and therefore that the conclusions of this paper apply to arbitrary physical states of the theory. This also ties in with the expectations from the symmetric orbifold.} One expects the $\tilde{W}_{\Gamma}$ to be constants, independent of $\Gamma$, but it would be good to deduce this  (as well as the complete form of (\ref{full-corr})) directly from solving the conformal bootstrap equations on the world-sheet.

\subsection{Relation to twistor string theory}\label{twstr}

We have seen above how the relation \eqref{eq:magic} is central to the unusual property of the physical correlators being delta function localised to those Riemann surfaces which admit a finite branched covering $x=\Gamma(z)$ of the boundary sphere. The relation  \eqref{eq:magic} actually also gives a more geometric picture of our free field realisation of the $k=1$ sigma model on ${\rm AdS}_3\times {\rm S}^3$. It directly implies that the symplectic bosons $\xi^{\pm}(z)$ that appear in \eqref{eq:magic} are to be thought of as homogeneous coordinates in the complex parametrisation of the target $\mathbb{CP}^1$ on the boundary of ${\rm AdS}_3$. 

Note that in the conventional NS-R parametrisation of the ${\rm AdS}_3$ sigma model which was employed in \cite{Eberhardt:2019ywk}, the field $\gamma(z)$ (and its complex conjugate) parametrise the sphere direction. In that case we argued, based on the Ward identities, that $\gamma(z)=\Gamma(z)$ holds as a relation within physical correlators (see eq.~(6.4) of \cite{Eberhardt:2019ywk}). In the free field realisation, we do not need to go through the 
Ward identities to arrive at the underlying geometric picture of the path integral localising on  branched coverings. 

Furthermore, \eqref{eq:magic} is very suggestive of a (quantum sigma model version of) a twistorial incidence relation. Recall that in four dimensional (complexified) Minkowski space the fundamental twistor relation is
\begin{equation}
\mu_{\dot{a}}+x_{a\dot{a}}\lambda^{a}=0 \ ,
\label{twstr-rel}
\end{equation}
where the Minkowski coordinates $x_{a\dot{a}}=\sigma^{\mu}_{a\dot{a}}x_{\mu}$ are written in bispinor form, and $(\mu_{\dot{a}}, \lambda_{a})$ are the (bosonic) spinor variables which are homogeneous coordinates for the corresponding twistor space ($\mathbb{CP}^3$ in this case). 

In our case, $\xi^{\pm}(z)$ therefore play the role of twistor variables that coordinatise 
the $\mathbb{CP}^1$. That this is not an accident can be seen by looking at the free sigma model of the symplectic boson and fermions parametrising the $\mathfrak{psu}(1,1|2)_1$ theory. This can be cast in a form which is a two dimensional closed string analogue of the twistor open string theory proposed by Berkovits \cite{Berkovits:2004hg} for 4d super Yang-Mills. 
To see this, we define the supertwistor fields
\be
Z^I \equiv \bigl(\xi^+, \xi^-, \psi^+, \psi^-\bigr) \ , \qquad Y_I \equiv \bigl(- \eta^-, \eta^+, - \chi^-,\chi^+ \bigr) \ . 
\ee
The free field action splits into left and right moving parts which can each be written as 
\begin{eqnarray}
S_{\rm left} & = & \int d^2z \, Y_{LI} \nabla Z_L^I \nonumber \\
& = &  \int d^2z \, \Bigl( - \eta^- \bigl( \bar{\partial} - A \bigr) \, \xi^+ 
+  \eta^+ \bigl( \bar{\partial} -  A \bigr) \,  \xi^- 
- \chi^- \bigl( \bar{\partial} -  A \bigr) \, \psi^+ 
+ \chi^+ \bigl( \bar{\partial} -  A \bigr) \, \psi^- \Bigr) \nonumber \\
& = & \int d^2z \, \Bigl[ \bigl( - \eta^- \bar{\partial} \xi^+ + \eta^+ \bar{\partial} \xi^- 
- \chi^- \bar{\partial}\psi^+ + \chi^+ \bar{\partial}\psi^- \bigr) + J  \, A \Bigr] \ , 
\label{eq:twstr-act}
\end{eqnarray}
and similarly for the right moving piece. 
Here we have also introduced a $U(1)$ gauge field $A=A_{\bar{z}}$ under which the $Z^I$ fields carry charge $-1$ while the $Y_I$ fields carry charge $+1$. 
The corresponding left moving current is then 
\be
J = Y_I Z^I =  \eta^- \xi^+ - \eta^+ \xi^- + \chi^- \psi^+ - \chi^+ \psi^- =  2 U + 2 V =  2 Z \ .
\label{eq:twstr-curr}
\ee
In other words, this is nothing other than the gauging of the $Z$ current (not to be confused with the supertwistor $Z^I$!), which is necessary in order to describe the $\mathfrak{psu}(1,1|2)$ model, see the discussion below eq.~(\ref{ZZ}). 
In this form, the action \eqref{eq:twstr-act} and the current \eqref{eq:twstr-curr} are the direct analogues of eqs.~(6) and (8) of \cite{Berkovits:2004hg}, respectively. 

{}From this perspective it thus seems natural to view $Z^I$ as homogeneous coordinates for the supertwistor space 
$\mathbb{CP}^{1|2}$ --- the lower dimensional analogue to the $\mathbb{CP}^{3|4}$ that enters in the usual twistor string theory. Thus \eqref{eq:magic} is the natural bosonic incidence relation for these variables. Also recall that the twistor string theory of Berkovits \cite{Berkovits:2004hg} is a topologically twisted string theory which had physical correlators localised on the loci of incidence relations such as these, see in particular eq.~(16) of \cite{Berkovits:2004hg}.\footnote{A similar statement also applies to the twistor string theory of Witten \cite{Witten:2003nn}. See for instance, eq.~(1) of \cite{Berkovits:2004hg} or eq.~(1.1) of \cite{Roiban:2004yf}. However, in that context the $\delta$-function localisation happens on the worldvolume of a D-string.} This again fits with the expectation that there is an underlying topological string description of the $\mathfrak{psu}(1,1|2)_1$ sigma model as well. This is currently under investigation \cite{GG}.

\section{Explicit calculations}\label{sec:examples}

In Section \ref{sec:twistor} we used general arguments from complex analysis to derive a relationship between the correlation functions \eqref{correlators1} and the covering map. The relation in equation \eqref{eq:magic}, along with its derivation, allowed us to impose strong constraints on the form of the correlation functions. Namely, we were able to show that the correlator \eqref{correlators1} is delta-function localised to the insertion points that allow for the existence of a covering map. Furthermore, using \eqref{eq:magic} we could fully constrain the dependence of the correlation functions on the $J^3_0$ eigenvalue, see eq.~\eqref{general-correlator}.

As we mentioned before, we could also determine the correlators using a more direct approach, following in essence the techniques of \cite{Eberhardt:2019ywk}, but now working directly with the symplectic boson fields (rather than the currents in $\mathfrak{sl}(2,\mathds{R})$). 
That is, we could constrain the correlation functions of our theory via the local Ward identities that arise upon inserting $\xi^{\pm}(z)$ and $\eta^{\pm}(z)$ into a correlation function. In this section, we spell out the details of this approach, and demonstrate that it can be used to exemplify the results of Section \ref{sec:twistor} in a few simple examples.

Note that the following discussion will be done for the free field theory corresponding to $\mathfrak{u}(1,1|2)$; in order to obtain the results relevant for $\mathfrak{psu}(1,1|2)$ we have to divide out the $Z_n$ generators as discussed below eq.~(\ref{ZZ}). Since we will be considering correlators of vertex operators that are primary with respect to $Z_n$ (and do not involve any $Z_{-n}$ descendants with $n>0$) there is essentially no difference between the two calculations; the only small modification one has to take into account is  that some of our fields carry non-trivial $U_0$-charge, and that this gives rise to an additional contribution of the form 
\be\label{ufactor}
\langle V_{q_1}(z_1) \cdots V_{q_n} (z_n) \rangle = \prod_{i<j} (z_i - z_j)^{- 2 q_i q_j} \ ,
\ee
where $V_q(z)$ is the vertex operator corresponding to the primary state with $U_0$ charge $U_0=q$. In particular, in the physical correlators in \eqref{correlators1}, we will get additional such contributions from the $W$ fields (that have $U_0$ charge equal to $q=1$) as well as $(n-2)$ of the $V^{w}_{m_1, m_2}(x; z)$, see eq.~(\ref{correlators1}). These contributions will cancel against the explicit dependence on the $u_{\alpha}$ that we will find below. This is how it has to be since in the $\mathfrak{psu}(1,1|2)$ theory, the $W$ field describes the vacuum, and hence does not contribute to the correlators.

\subsection{Ward identities}\label{sec:Ward}

The local Ward identities for the free field realisation of the $\mathfrak{psu}(1,1|2)_1$ model can be deduced by considering the insertion of $\xi^{\pm}(z)$ and $\eta^{\pm}(z)$ into correlation functions. We begin with the correlators
\begin{equation}
\Bigl\langle\xi^{\pm}(z)\prod_{\alpha=1}^{n-2}W(u_{\alpha})\prod_{i=1}^{n}V_{m_1^i,m_2^i}^{w_i}(x_i;z_i)\Bigr\rangle\ .
\label{eq:xi-inserted}
\end{equation}
For the time being, let us assume that all of the $w_i$ are odd, so that \eqref{eq:xi-inserted} defines a meromorphic function of $z$ with no branch cuts. (The strategy for even spectral flow is demonstrated in Appendix \ref{app:explicit}.) As a function of $z$, \eqref{eq:xi-inserted} has poles at $z=z_i$ with order $\frac{w_i+1}{2}$, and behaves asymptotically as $\mathcal{O}(1/z)$ as $z\to\infty$. Thus, by Liouville's theorem, we can construct the correlator knowing only the Laurent expansion near its poles. Using the OPEs \eqref{OPE2} and \eqref{OPE4}, as well as \eqref{eq:xi-shifted}, it therefore follows that 
\begin{equation}
\begin{split}
\Bigl\langle\xi^{+}(z)\prod_{\alpha=1}^{n-2}W(u_{\alpha})\prod_{i=1}^{n}V_{m_1^i,m_2^i}^{w_i}(x_i;z_i)\Bigr\rangle&=\sum_{i=1}^{n}\sum_{\ell=1}^{\frac{w_i+1}{2}}\frac{1}{(z-z_i)^{\ell}}F^{i}_{\ell-\frac{1}{2}}\ ,\\
\Bigl\langle\xi^{-}(z)\prod_{\alpha=1}^{n-2}W(u_{\alpha})\prod_{i=1}^{n}V_{m_1^i,m_2^i}^{w_i}(x_i;z_i)\Bigr\rangle&=-\sum_{i=1}^{n}\sum_{\ell=1}^{\frac{w_i+1}{2}}\frac{x_i}{(z-z_i)^{\ell}}F^{i}_{\ell-\frac{1}{2}}\ .
\end{split}
\label{eq:xi-local-ward}
\end{equation}
Here, we have defined
\begin{equation}
F^i_{r}=\Bigl\langle\prod_{\alpha=1}^{n-2}W(u_{\alpha})\left(\xi^+_{r}V_{m_1^i,m_2^i}^{w_i}(x_i;z_i)\right)\prod_{j\neq i}^{n}V^{w_j}_{m_1^j,m_2^j}(x_j;z_j)\Bigr\rangle\ .
\end{equation}

The Ward identities \eqref{eq:xi-local-ward} involve several `unknown' correlation functions that we would like to eliminate. This is done by noticing that, since $\xi^{-(x)}(z)+x\, \xi^{+(x)}(z)=\xi^-(z)$, see eq.~\eqref{eq:xi-shifted}, the OPE of $\xi^-(z)+x_k\, \xi^+(z)$ with $V_{m_1^k,m_2^k}^{w_k}(x_k;z_k)$ has a zero of order $\frac{w_k-1}{2}$. That is,
\begin{equation}
\Bigl\langle\left(\xi^-(z)+x_k\, \xi^+(z)\right)\prod_{\alpha=1}^{n-2}W(u_{\alpha})\prod_{i=1}^{n}V_{m_1^i,m_2^i}^{w_i}(x_i;z_i)\Bigr\rangle\sim\mathcal{O}\left((z-z_k)^{\frac{w_k-1}{2}}\right)\ ,
\end{equation}
as $z\to z_k$. Using \eqref{eq:xi-local-ward}, we can write this as
\begin{equation}
\label{eq:xi-regularity-constraint}
\sum_{i\neq k}\sum_{\ell=1}^{\frac{w_i+1}{2}}\frac{x_k-x_i}{(z-z_i)^{\ell}}F^i_{\ell-\frac{1}{2}}\sim\mathcal{O}\left((z-z_k)^{\frac{w_k-1}{2}}\right),\quad z\to z_k\ .
\end{equation}
Furthermore, \eqref{OPE6} tells us that the OPE of $\xi^{\pm}(z)$ with $W(u_{\alpha})$ has a zero of order 1 at $z=u_{\alpha}$, i.e.\ 
\begin{equation}
\Bigl\langle\xi^\pm(z)\prod_{\alpha=1}^{n-2}W(u_{\alpha})\prod_{i=1}^{n}V_{m_1^i,m_2^i}^{w_i}(x_i;z_i)\Bigr\rangle\sim\mathcal{O}\left((z-u_{\alpha})\right),\quad z\to u_{\alpha}\ .
\end{equation}
Using \eqref{eq:xi-local-ward}, we can thus write this as 
\begin{equation}
\sum_{i=1}^{n}\sum_{\ell=1}^{\frac{w_i+1}{2}}\frac{1}{(u_{\alpha}-z_i)^{\ell}}F^i_{\ell-\frac{1}{2}}=\sum_{i=1}^{n}\sum_{\ell=1}^{\frac{w_i+1}{2}}\frac{x_i}{(u_{\alpha}-z_i)^{\ell}}F^i_{\ell-\frac{1}{2}}=0\ .
\label{eq:ualpha-constraint}
\end{equation}

Equations \eqref{eq:xi-regularity-constraint} and \eqref{eq:ualpha-constraint} form a highly constraining homogeneous linear system for the correlators $F^i_{r}$. There are $\sum_{i=1}^{n}\tfrac{w_i+1}{2}=N+n-1$ such correlators, and we have $2(n-2)+\sum_{i=1}^{n}\tfrac{w_i-1}{2}=N+2(n-2)-1$ constraints, where $N$ is given by \eqref{eq:Riemann-Hurwitz}. Thus, solutions to this system only exist given that $n-3$ conditions on the parameters $(x_i,z_i,u_{\alpha})$ are satisfied. It will turn out, as we shall see, that the existence of a solution is independent of $u_{\alpha}$, and the appropriate conditions for a solution to exist will be those required for the existence of a covering map.

\subsubsection{The recursion relations}

Equations \eqref{eq:xi-regularity-constraint} and \eqref{eq:ualpha-constraint} on their own do not give us a great deal of information, since they relate the unknown correlators $F^i_{r}$ to each other. However, since 
\be
\xi^+_{\frac{w}{2}}[|m_1,m_2\rangle]^{\sigma^w}=[|m_1,m_2+\tfrac{1}{2}\rangle]^{\sigma^w}\ ,
\ee
we can relate the unknown $F^i_{\frac{w_i}{2}}$ to the original correlation function as
\begin{equation}
F^{i}_{\frac{w_i}{2}}=\Bigl\langle\prod_{\alpha=1}^{n-2}W(u_{\alpha})\left(V^{w_i}_{m_1^i,m_2^i+\frac{1}{2}}\right)(x_i;z_i)\prod_{j\neq i}^{n}V^{w_j}_{m_1^j,m_2^j}(x_j;z_j)\Bigr\rangle=\bigl\langle m_2^i+\tfrac{1}{2}\bigr\rangle\ ,
\end{equation}
where we have introduced the shorthand notation $\langle m_2^i+\frac{1}{2}\rangle$ to represent the correlator \eqref{correlators1} with the shift $m_2^i\to m_2^i+\frac{1}{2}$. Futhermore, using \eqref{OPE4}, we can determine the coefficient of the $\mathcal{O}((z-z_k)^{\frac{w_k-1}{2}})$ term in \eqref{eq:xi-regularity-constraint} to be
\begin{equation}
\Bigl\langle\prod_{\alpha=1}^{n-2}W(u_{\alpha})\left(\xi^{-}_{-\frac{w_k}{2}}V^{w_k}_{m_1^k,m_2^k}\right)(x_k;z_k)\prod_{j\neq k}^{n}V^{w_j}_{m_1^j,m_2^j}(x_j;z_j)\Bigr\rangle=-\bigl\langle m_1^k-\tfrac{1}{2}\bigr\rangle\ ,
\end{equation}
where we have used the fact that $\xi^-_{-\frac{w}{2}}[ | m_1,m_2 \rangle]^{\sigma^w}=-[| m_1-\tfrac{1}{2},m_2\rangle]^{\sigma^w}$, and we have again used the shorthand in which $\langle m_1^k-\tfrac{1}{2}\rangle$ denotes the correlator with the shift $m_1^k\to m_1^k-\frac{1}{2}$. This allows us to write \eqref{eq:xi-regularity-constraint} as a linear system given by
\begin{equation}
\begin{split}
\sum_{i\neq k}\frac{x_k-x_i}{(z_k-z_i)^{\frac{w_i+1}{2}+p}}\binom{\frac{w_i-1}{2}+p}{p}\bigl\langle m_2^i+\tfrac{1}{2}\bigr\rangle+\sum_{i\neq k}\sum_{\ell=1}^{\frac{w_i-1}{2}}&\frac{x_k-x_i}{(z_k-z_i)^{\ell+p}}\binom{\ell+p-1}{p}F^{i}_{\ell-\frac{1}{2}}\\
&=(-1)^{p+1}\bigl\langle m_1^k-\tfrac{1}{2}\bigr\rangle\delta_{p,\frac{w_i-1}{2}}\ ,
\end{split}
\label{eq:xi-linear-system-1}
\end{equation}
where $p\in\{0,\ldots,\frac{w_k-1}{2}\}$, while \eqref{eq:ualpha-constraint} leads to 
\begin{equation}
\begin{split}
\sum_{i=1}^{n}\frac{\bigl\langle m_2^i+\tfrac{1}{2}\bigr\rangle}{(u_{\alpha}-z_i)^{\frac{w_i+1}{2}}}+\sum_{i=1}^{n}\sum_{\ell=1}^{\frac{w_i-1}{2}}\frac{F^i_{\ell-\frac{1}{2}}}{(u_{\alpha}-z_i)^{\ell}}&=0\ ,\\
\sum_{i=1}^{n}\frac{x_i\bigl\langle m_2^i+\tfrac{1}{2}\bigr\rangle}{(u_{\alpha}-z_i)^{\frac{w_i+1}{2}}}+\sum_{i=1}^{n}\sum_{\ell=1}^{\frac{w_i-1}{2}}\frac{x_i F^i_{\ell-\frac{1}{2}}}{(u_{\alpha}-z_i)^{\ell}}&=0\ .
\end{split}
\label{eq:xi-linear-system-2}
\end{equation}
Equations \eqref{eq:xi-linear-system-1} and \eqref{eq:xi-linear-system-2} can be used to eliminate the unknown $F^i_{r-1/2}$ correlators with $r\leq\frac{w_i-1}{2}$, and the remaining  constraints define a system of recursion relations in the variables $m_1^i,m_2^i$ for the correlation functions \eqref{correlators1}. These relations, when solved, reproduce the structure \eqref{general-correlator}, as we will see in the examples below.

\subsubsection[The \texorpdfstring{$\eta^{\pm}$}{eta} constraints]{\boldmath The \texorpdfstring{$\eta^{\pm}$}{eta} constraints}\label{etaconstraints}

So far we have only concerned ourselves with the constraint equations coming from the insertion of the fields $\xi^{\pm}$ into correlation functions. As it turns out, the constraints coming from the $\xi^{\pm}$ insertions are both technically simpler than those for the insertion of $\eta^{\pm}$ and constrain the form of the correlation functions almost fully. The $\eta^{\pm}$ constraints are, however, important for deriving the $U_0$ charge constraint discussed in Section \ref{sec:U0charge}, and thus we present them here.

Consider the correlation function
\begin{equation}
\Bigl\langle\eta^{\pm}(z)\prod_{\alpha=1}^{n-2}W(u_{\alpha})\prod_{i=1}^{n}V^{w_i}_{m_1^i,m_2^i}(x_i;z_i)\Bigr\rangle \ .
\end{equation}
Just as with the $\xi^{\pm}$ insertions, we can use the OPE of $\eta^{\pm}$ with the vertex operators to constrain the form of this correlator. This is slightly more complicated than for the $\xi^{\pm}$ insertions, since the OPE of $\eta^{\pm}$ with $W$ has a first order pole. The result is
\begin{equation}
\begin{split}
\Bigl\langle\eta^+(z)\prod_{\alpha=1}^{n-2}W(u_{\alpha})\prod_{i=1}^{n}V^{w_i}_{m_1^i,m_2^i}(x_i;z_i)\Bigr\rangle&=\sum_{i=1}^{n}\sum_{\ell=1}^{\frac{w_i+1}{2}}\frac{1}{(z-z_i)^{\ell}}\widetilde{F}^i_{\ell-\frac{1}{2}}+\sum_{\alpha=1}^{n-2}\frac{1}{z-u_{\alpha}}A^+_{\alpha}\ ,\\
\Bigl\langle\eta^-(z)\prod_{\alpha=1}^{n-2}W(u_{\alpha})\prod_{i=1}^{n}V^{w_i}_{m_1^i,m_2^i}(x_i;z_i)\Bigr\rangle&=-\sum_{i=1}^{n}\sum_{\ell=1}^{\frac{w_i+1}{2}}\frac{x_i}{(z-z_i)^{\ell}}\widetilde{F}^i_{\ell-\frac{1}{2}}+\sum_{\alpha=1}^{n-2}\frac{1}{z-u_{\alpha}}A^-_{\alpha}\ ,
\end{split}
\end{equation}
where we have defined
\begin{equation}
\begin{split}
\widetilde{F}^i_{r}&=\Bigl\langle\prod_{\alpha=1}^{n-2}W(u_{\alpha})\left(\eta^+_{r}V_{m_1^i,m_2^i}^{w_i}(x_i;z_i)\right)\prod_{j\neq i}^{n}V^{w_j}_{m_1^j,m_2^j}(x_j;z_j)\Bigr\rangle\ ,\\
A^{\pm}_{\alpha}&=\Bigl\langle\Bigl(\eta^{\pm}_{\frac{1}{2}}W\bigr)(u_{\alpha})\prod_{\beta\neq\alpha}^{n-2}W(u_{\beta})\prod_{i=1}^{n}V^{w_i}_{m_1^i,m_2^i}(x_i;z_i)\Bigr\rangle\ .
\end{split}
\end{equation}
Just as for the $\xi^{\pm}$ analysis, the linear combination $\eta^-(z)+x_k\, \eta^+(z)$ has a highly regular OPE with $V^{w_k}_{m_1^k,m_2^k}(x_k;z_k)$. That is,
\begin{equation}\label{etareg}
\Bigl\langle\left(\eta^-(z)+x_k\eta^+(z)\right)\prod_{\alpha=1}^{n-2}W(u_{\alpha})\prod_{i=1}^{n}V^{w_i}_{m_1^i,m_2^i}(x_i;z_i)\Bigr\rangle\sim\mathcal{O}\left((z-z_k)^{\frac{w_k-1}{2}}\right),\quad z\to z_k\ .
\end{equation}
Explicitly, this imposes the constraints
\begin{equation}
\sum_{i\neq k}^{n}\sum_{\ell=1}^{\frac{w_i+1}{2}}\frac{x_k-x_i}{(z-z_i)^{\ell}}\widetilde{F}^i_{\ell-\frac{1}{2}}+\sum_{\alpha=1}^{n-2}\frac{A^-_{\alpha}+x_kA^+_{\alpha}}{z-u_{\alpha}}\sim\mathcal{O}\left((z-z_k)^{\frac{w_k-1}{2}}\right),\quad z\to z_k\ .
\label{eq:eta-regularity-constraints}
\end{equation}
Unlike the $\xi^{\pm}$ case, however, there are no constraints coming from the behavior of this correlator near $z=u_{\alpha}$. Just as before, since we know the action of $\eta^+_{\frac{w}{2}}$ and $\eta^-_{-\frac{w}{2}}$ on the state $[|m_1,m_2\rangle]^{\sigma^w}$, we also know the form of $\widetilde{F}^{i}_{\frac{w_i}{2}}$, as well as the lowest-order coefficient on the right-hand side of \eqref{eq:eta-regularity-constraints}. Putting this together, we end up with the linear system
\begin{equation}
\begin{split}
\sum_{i\neq k}^{n}\frac{x_k-x_i}{(z_k-z_i)^{\frac{w_i+1}{2}+p}}\binom{\frac{w_i-1}{2}+p}{p}2m_1^i\bigl\langle m_1^i+\tfrac{1}{2}\bigr\rangle+\sum_{i\neq k}^{n}\sum_{\ell=1}^{\frac{w_i-1}{2}}\frac{x_k-x_i}{(z_k-z_i)^{\ell+p}}\binom{\ell+p-1}{p}\widetilde{F}^i_{\ell-\frac{1}{2}}\\
+\sum_{\alpha=1}^{n-2}\frac{A^-_{\alpha}+x_kA^+_{\alpha}}{(z_k-u_{\alpha})^{p+1}}=(-1)^{p+1}2m_2^k\bigl\langle m_2^k-\tfrac{1}{2}\bigr\rangle\delta_{p,\frac{w_k-1}{2}}
\end{split}
\label{eq:eta-linear-system}
\end{equation}
for $p=0,\ldots,\frac{w_k-1}{2}$. Equation \eqref{eq:eta-linear-system} defines a linear system of $\sum_{i=1}^{n}\frac{w_i+1}{2}=N+n-1$ equations for $2(n-2)+\sum_{i=1}^{n}\frac{w_i-1}{2}=N+2n-5$ unknowns, and is thus under-constrained for $n>4$.

\subsection{Some three-point functions}\label{sec:3pt-111}

Let us begin by demonstrating the utility of the local Ward identities from above in the simplest possible case: the three-point function with unit spectral flow. That is, we aim to constrain the correlator
\begin{equation}
\Bigl\langle W(u)\prod_{i=1}^{3}V^1_{m_1^i,m_2^i}(x_i;z_i)\Bigr\rangle\ .
\end{equation}
Let us start with the $\xi^{\pm}$ constraints. In this simple case, the linear system defined by \eqref{eq:xi-linear-system-1} and \eqref{eq:xi-linear-system-2} takes the form
\begin{equation}
\begin{split}
\sum_{i\neq k}^{3}\frac{x_{k}-x_i}{z_k-z_i}\bigl\langle m_2^i+\tfrac{1}{2}\bigr\rangle=-\bigl\langle m_1^k-\tfrac{1}{2}\bigr\rangle\ ,\quad\sum_{i=1}^{3}\frac{\bigl\langle m_1^i-\tfrac{1}{2}\bigr\rangle}{u-z_i}=0\ ,\quad\sum_{i=1}^{3}\frac{x_i\bigl\langle m_1^i-\tfrac{1}{2}\bigr\rangle}{u-z_i}=0\ .
\end{split}
\label{eq:111-linear-system}
\end{equation}
By the global $\mathfrak{sl}(2,\mathbb{R})$ symmetry of the theory, we can perform a M\"obius transformation to bring $x_1=z_1=0$, $x_2=z_2=1$, and $x_3=z_3\to\infty$. With this choice, the first equation in \eqref{eq:111-linear-system} becomes
\begin{equation}
\sum_{i\neq k}^{3}\bigl\langle m_2^i+\tfrac{1}{2}\bigr\rangle+\bigl\langle m_1^k-\tfrac{1}{2}\bigr\rangle=0\ ,
\label{eq:111-regularity-constraint-simplified}
\end{equation}
while the last two are easily solved to yield
\begin{equation}
\bigl\langle m_1^2-\tfrac{1}{2}\bigr\rangle=-\frac{u-1}{u}\bigl\langle m_1^1-\tfrac{1}{2}\bigr\rangle\ ,\qquad\qquad  \bigl\langle m_1^3-\tfrac{1}{2}\bigr\rangle=-\frac{1}{u}\bigl\langle m_1^1-\tfrac{1}{2}\bigr\rangle\ .
\label{eq:111-u-constraint-simplified}
\end{equation}
Together, \eqref{eq:111-u-constraint-simplified} and \eqref{eq:111-regularity-constraint-simplified} imply that
\begin{equation}
\bigl\langle m_1^i+\tfrac{1}{2}\bigr\rangle=\bigl\langle m_2^i-\tfrac{1}{2}\bigr\rangle
\label{eq:111-simple-recursion}
\end{equation}
for all $i$. The most general solution to \eqref{eq:111-simple-recursion} and \eqref{eq:111-u-constraint-simplified} is given by
\begin{equation}
\begin{split}
\Bigl\langle W(u)V^1_{m_1^1,m_2^1}(0;0)V^1_{m_1^2,m_2^2}(1;1)&V^1_{m_1^3,m_2^3}(\infty;\infty)\Bigr\rangle\\
&=C(u,\mathcal{S})(-u)^{-2(m_1^1-m_2^1)}(u-1)^{-2(m_1^2-m_2^2)}\ ,
\end{split}
\label{eq:111-solution}
\end{equation}
where $C(u,\mathcal{S})$ is an arbitrary function of $u$\footnote{$C(u,\mathcal{S})$ could also depend on the $m_j^i$ mod $\frac{1}{2}$.} and
\begin{equation}
\mathcal{S}:=m_1^1-m_2^1+m_1^2-m_2^2+m_1^3-m_2^3\ .
\end{equation}
The $u$-dependence of eq.~(\ref{eq:111-solution}) is an artefact of the fact that we are working with the full free field $\mathfrak{u}(1,1|2)$ theory, rather than with $\mathfrak{psu}(1,1|2)_1$. In particular, the $u$-dependence of  eq.~(\ref{eq:111-solution}) disappears upon removing the factor that comes from the $U_0$ charge terms, see eq.~(\ref{ufactor}) and the discussion below. 

Note that eq.~\eqref{eq:111-solution} matches the analysis of Section~\ref{sec:twistor}, particularly eq.~\eqref{general-correlator}. Indeed, the relevant covering map in this case is  simply $\Gamma(z)=z$. Since $\Gamma(z)=z_i+(z-z_i)$, the constant $a_i^{\Gamma}$ in \eqref{general-correlator} is always just $1$. Thus, we should expect the three-point function with unit spectral flow to have no dependence on $h_i:=m_1^i+m_2^i+\frac{1}{2}$, and this is indeed what we have found.

Now, let us consider the $\eta^{\pm}$ constraints for this correlator. With our choice of $x_i,z_i$, \eqref{eq:eta-linear-system} becomes
\begin{equation}
\sum_{i\neq k}2m_1^i\bigl\langle m_1^i+\tfrac{1}{2}\bigr\rangle+\frac{A^-+x_kA^+}{z_k-u}=-2m_2^k\bigl\langle m_2^k-\tfrac{1}{2}\bigr\rangle\ .
\end{equation}
This is a system of three equations with two unknowns, $A^{\pm}$, and thus the unknowns can be eliminated. After eliminating $A^{\pm}$ and using the solution \eqref{eq:111-solution}, we are left with the simple constraint
\begin{equation}
C(u,\mathcal{S}+\tfrac{1}{2})\, \mathcal{S}=0\ .
\end{equation}
Thus, $C(u,\mathcal{S})$ must be proportional to $C(u)\delta_{\mathcal{S},\tfrac{1}{2}}$ for some function $C(u)$, and the three-point function takes the form
\begin{equation}
\begin{split}
\Bigl\langle W(u)V^1_{m_1^1,m_2^1}(0;0)V^1_{m_1^2,m_2^2}(1;1)&V^1_{m_1^3,m_2^3}(\infty;\infty)\Bigr\rangle\\
&=C(u)(-u)^{-2(m_1^1-m_2^1)}(u-1)^{-2(m_1^2-m_2^2)}\delta_{\mathcal{S},\tfrac{1}{2}}\ .
\end{split}
\label{eq:111-solution-final}
\end{equation}
It is worth noting that the constraint $\mathcal{S}=\frac{1}{2}$ is exactly the statement of the $U_0$ charge condition in Section \ref{sec:U0charge}. We should also mention that, because of (\ref{ufactor}), we know that the function $C(u)$ must be of the form $C(u) = C u (u-1)$, so that the total $u$-dependence has the expected form. (Recall that the $U_0$-charge of the vertex operator $V^1_{m_1,m_2}$ is $q = m_1 -m_2 - \frac{1}{2}$.)

Finally, let us check that \eqref{eq:magic} actually holds for this three-point function. The appropriate covering map is simply $\Gamma(z)=z$. Using the local Ward identities and the solution \eqref{eq:111-solution-final}, we have
\begin{equation}
\Bigl\langle\left(\xi^-(z)+\Gamma(z)\,\xi^+(z)\right)W(u)\prod_{i=1}^{3}V_{m_1^i,m_2^i}^{w_i}(x_i;z_i)\Bigr\rangle=\sum_{i=1}^{3}\bigl\langle m_2^i+\tfrac{1}{2}\bigr\rangle=0\ .
\end{equation}

\subsubsection[The three-point function with \texorpdfstring{$(w_1,w_2, w_3)=(3,3,1)$}{w = (3,3,1)}]{\boldmath The three-point function with \texorpdfstring{$(w_1, w_2, w_3)=(3,3,1)$}{w = (3,3,1)}}
\label{sec:3pt-331}

The computation of three-point functions with higher values of the spectral flow parameters $(w_1, w_2, w_3)$ works similarly to the $(w_1, w_2, w_3)=(1,1,1)$ case discussed above.  Let us summarise the main steps of the computation by means of an example, the three-point function 
\begin{equation}
\Bigl\langle W(u) \, V^3_{m_1^1,m_2^1}(0;0) \, V^3_{m_1^2,m_2^2}(1;1) \,  V^1_{m_1^3,m_2^3}(\infty;\infty) \Bigr\rangle\ .
\end{equation}
Eqs.~\eqref{eq:xi-linear-system-1} and \eqref{eq:xi-linear-system-2} define a linear system of seven equations in the seven unknowns 
\be
F^1_{\frac{1}{2}} \ , \quad F^2_{\frac{1}{2}} \ , \quad \bigl\langle m_1^i-\tfrac{1}{2}\bigr\rangle \text{ with } i=1,2,3 \ , \quad \bigl\langle m_2^j+\tfrac{1}{2}\bigr\rangle \text{ with } j=2,3 \ . 
\ee 
The solution of the linear system implies
\begin{subequations}
\begin{align}
\bigl\langle m_1^1-\tfrac{1}{2}\bigr\rangle & = \bigl\langle m_2^1+\tfrac{1}{2}\bigr\rangle \ , \\
\bigl\langle m_1^2-\tfrac{1}{2}\bigr\rangle & = \bigl\langle m_2^2+\tfrac{1}{2}\bigr\rangle = \frac{u-1}{u} \, \bigl\langle m_2^1+\tfrac{1}{2}\bigr\rangle \ , \\
\bigl\langle m_1^3-\tfrac{1}{2}\bigr\rangle & = 3 \,  \bigl\langle m_2^3+\tfrac{1}{2}\bigr\rangle=\frac{3}{u} \, \bigl\langle m_2^1+\tfrac{1}{2}\bigr\rangle \ ,  
\end{align}
\end{subequations}
reproducing eq.~\eqref{magic-recursion} with 
\be 
a_1^\Gamma = 1 \ , \qquad a_2^\Gamma = 1 \ , \qquad a_3^\Gamma = 3 \ . 
\ee
The correlator is then fixed to be 
\begin{multline}
\Bigl\langle W(u) \, V^3_{m_1^1,m_2^1}(0;0) \, V^3_{m_1^2,m_2^2}(1;1) \,  V^1_{m_1^3,m_2^3}(\infty;\infty) \Bigr\rangle \\
= C(u) \, 3^{-(m_1^3+m_2^3)} \, \left(\frac{3}{u^2} \right)^{m_1^1-m_2^1} \, \left( \frac{3}{(u-1)^2} \right)^{m_1^2-m_2^2} \, \delta_{\mathcal S,\frac{1}{2}} \ ,  
\label{eq:explicit-correlator-331}
\end{multline}
where again $C(u) = C u (u-1)$, see the discussion below eq.~(\ref{eq:111-solution-final}). The constraint $\mathcal{S}=\frac{1}{2}$ follows again from the $\eta^\pm$ constraints \eqref{eq:eta-linear-system}. Finally, one easily checks that the incidence relation \eqref{eq:magic} is indeed obeyed. 

We have performed a similar analysis for all three-point functions with $w_i$ odd and $w_i \leq 19$, and shown that it leads to the correct covering map solution (provided that the selection rules (\ref{eq:fusion-rules}) are satisfied). We have also checked that this  solution indeed satisfies the incidence relation for $w_i\leq 9$.

\subsection[The four-point function with \texorpdfstring{$(w_1, w_2, w_3, w_4)=(1,1,1,1)$}{w = (1,1,1,1)}]{\boldmath The four-point function with \texorpdfstring{$(w_1, w_2, w_3, w_4)=(1,1,1,1)$}{w = (1,1,1,1)}}

The three-point function examples of the previous subsection demonstrated several key features of correlation functions in this model, including the identity \eqref{eq:magic}. However, as a covering map always exists for three-point functions,  one does not see the localisation of correlation functions at the level of the three-point function. To provide an explicit example of how localisation emerges for higher-point functions, we turn our attention to a four-point function. We will concentrate on the simplest case where all spectral flows are equal to one, i.e.
\begin{equation}
\Bigl\langle \prod_{\alpha=1}^{2}W(u_\alpha)\prod_{i=1}^{4}V_{m_1^i,m_2^i}^{1}(x_i;z_i)\Bigr\rangle \ .
\label{eq:4pt-1111}
\end{equation}

\subsubsection{Localisation}
The relevant covering map for the $w=1$ four-point function is simply given by a M\"obius transformation satisfying $\Gamma(z_i)=x_i$. Such a function only exists provided that the cross-ratios of the two sets of variables coincide, i.e.\ provided that 
\be\label{cross}
\frac{(z_1-z_2) (z_3 - z_4)}{(z_1 - z_3) (z_2 - z_4)} = \frac{(x_1-x_2) (x_3 - x_4)}{(x_1 - x_3) (x_2 - x_4)} \ . 
\ee
We want to show in the following that the correlation function indeed localises to this configuration, i.e.\ that it vanishes unless (\ref{cross}) is satisfied. Adopting the short-hand notation of the previous section, eqs.~(\ref{eq:xi-linear-system-1}) and (\ref{eq:xi-linear-system-2}) take the form 
\begin{equation}
\sum_{i=1}^{4}\frac{1}{u_{\alpha}-z_i}\bigl\langle m_2^i+\tfrac{1}{2} \bigr\rangle
=\sum_{i=1}^{4}\frac{x_i}{u_{\alpha}-z_i}\bigl\langle m_2^i+\tfrac{1}{2} \bigr\rangle=0\ ,\qquad \alpha=1,2\ .
\label{eq:1111-u-regularity-1}
\end{equation}
\noindent Just as for the three-point function, we can use the conformal Ward identities to set 
$x_1=z_1=0$, $x_2=z_2=1$, and $x_3=z_3=\infty$, in which case the relevant covering map is simply $\Gamma(z)=z$. Equation \eqref{eq:1111-u-regularity-1} becomes then
\begin{subequations}
\begin{align}
\frac{1}{u_{\alpha}} \bigl\langle m_2^1+\tfrac{1}{2} \bigr\rangle+\frac{1}{u_{\alpha}-1} \bigl\langle m_2^2+\tfrac{1}{2} \bigr\rangle +\frac{1}{u_{\alpha}-z_4}  \bigl\langle m_2^4+\tfrac{1}{2} 
\bigr\rangle &=0\ , \label{eq:1111-u-regularity-2a}
\\
\frac{1}{u_{\alpha} -1} \bigl\langle m_2^2+\tfrac{1}{2} \bigr\rangle -\bigl\langle m_2^3+\tfrac{1}{2} \bigr\rangle +\frac{x_4}{u_{\alpha}-z_4}  \bigl\langle m_2^4+\tfrac{1}{2} 
\bigr\rangle &=0 \ . \label{eq:1111-u-regularity-2b}
\end{align}
\label{eq:1111-u-regularity-2}
\end{subequations}
Multiplying \eqref{eq:1111-u-regularity-2a} by $u_{\alpha}$, and then subtracting \eqref{eq:1111-u-regularity-2b}, one finds
\begin{equation}
\bigl\langle m_2^1+\tfrac{1}{2} \bigr\rangle+\bigl\langle m_2^2+\tfrac{1}{2} \bigr\rangle +\bigl\langle m_2^3+\tfrac{1}{2} \bigr\rangle +\frac{u_{\alpha}-x_4}{u_{\alpha}-z_4}\, \bigl\langle m_2^4+\tfrac{1}{2} \bigr\rangle =0 \ ,\qquad\alpha=1,2\ .
\end{equation}
The first three terms are independent of $\alpha$, and one immediately deduces 
\begin{equation}
\frac{u_1-x_4}{u_1-z_4} \, \bigl\langle m_2^4+\tfrac{1}{2} \bigr\rangle =\frac{u_2-x_4}{u_2-z_4}\, \bigl\langle m_2^4+\tfrac{1}{2} \bigr\rangle 
\end{equation}
and hence
\be 
(x_4-z_4)\, \bigl\langle m_2^4+\tfrac{1}{2} \bigr\rangle =0 \ .
\label{eq:4pt-1111-localisation-explicit}
\ee
Thus, the correlation function is non-zero only if $x_4=z_4=\Gamma(z_4)$. We also note that this has precisely the form of (\ref{deltacon}), and hence implies that the correlator has the structure 
\be 
\Bigl\langle \prod_{\alpha=1}^{2}W(u_\alpha) V_{m_1^1,m_2^1}^{1}(0;0) V_{m_1^2,m_2^2}^{1}(1;1) V_{m_1^3,m_2^3}^{1}(\infty;\infty) V_{m_1^4,m_2^4}^{1}(x_4;z_4) \Bigr\rangle \propto  \delta(x_4-z_4) \ . 
\ee

\subsubsection{Explicit form of correlator and incidence relation}

We can actually be more specific about the form of the correlator. To this end we deduce from eq.~(\ref{eq:xi-linear-system-1})  that  
\be 
- \bigl\langle m_1^i - \tfrac{1}{2} \bigr\rangle = \sum_{j\neq i} \frac{x_i-x_j}{z_i-z_j} \bigl\langle m_2^j + \tfrac{1}{2} \bigr\rangle \ , \qquad i = 1, \dots, 4 \ . 
\label{eq:4pt-1111-zi-regularity}
\ee
For our choice of insertion points and using \eqref{eq:4pt-1111-localisation-explicit}, this becomes 
\begin{subequations}
\begin{align}
& \bigl\langle m_1^1 -\tfrac{1}{2} \bigr\rangle =  \bigl\langle m_2^1 + \tfrac{1}{2} \bigr\rangle \ ,  \\
& \bigl\langle m_1^2 -\tfrac{1}{2} \bigr\rangle = \bigl\langle m_2^2 + \tfrac{1}{2} \bigr\rangle =  -\frac{ x_4 (u_1-1) (u_2-1) }{u_1 u_2 (x_4-1)} \bigl\langle m_2^1 + \tfrac{1}{2} \bigr\rangle \ , \\
& \bigl\langle m_1^3 -\tfrac{1}{2} \bigr\rangle = \bigl\langle m_2^3 + \tfrac{1}{2} \bigr\rangle =  -\frac{ x_4}{u_1 u_2} \bigl\langle m_2^1 + \tfrac{1}{2} \bigr\rangle \ , \\
& \bigl\langle m_1^4 -\tfrac{1}{2} \bigr\rangle = \bigl\langle m_2^4 + \tfrac{1}{2} \bigr\rangle =  \frac{ (u_1-x_4)(u_2-x_4)}{u_1 u_2 (x_4-1)} \bigl\langle m_2^1 + \tfrac{1}{2} \bigr\rangle \ . 
\end{align}
\label{eq:4pt-1111-system-solved}
\end{subequations} 
This set of recursion relations fixes the $m_1^i$ and $m_2^i$ dependence of the correlator as 
\begin{multline}
\Bigl\langle \prod_{\alpha=1}^{2}W(u_\alpha) V_{m_1^1,m_2^1}^{1}(0;0) V_{m_1^2,m_2^2}^{1}(1;1) V_{m_1^3,m_2^3}^{1}(\infty;\infty) V_{m_1^4,m_2^4}^{1}(x_4;z_4) \Bigr\rangle \\ = C(u_1,u_2, x_4, \mathcal{S}) \left( -\frac{u_1 u_2}{x_4} \right)^{-2(m_1^1-m_2^1)}\left( \frac{(u_1-1) (u_2-1)}{x_4-1} \right)^{-2(m_1^2-m_2^2)}  \ \times \\ 
\times \   \left( -\frac{(u_1-x_4) (u_2-x_4)}{x_4(x_4-1)} \right)^{-2(m_1^4-m_2^4)} \delta(x_4-z_4) \ , 
\label{eq:4pt-1111-mij-dependence}
\end{multline}
where $C(u_1,u_2, x_4, \mathcal{S})$ is a function of $u_1$, $u_2$, $x_4$, and 
\be 
\mathcal{S} = \sum_{i=1}^4 (m_1^i-m_2^i) \ . 
\label{eq:4pt-1111-S}
\ee
Note that eq.~\eqref{eq:4pt-1111-mij-dependence} does not carry any dependence on the $J_0^3$ eigenvalue $m_1^i+m_2^i$, in agreement with \eqref{general-correlator}. In fact, for the choice of insertion points made above, the covering map is simply $\Gamma(z) = z$, and $a_i^\Gamma = 1$ for $i=1, \dots, 4$. We also note that the $u_i$ dependence of eq.~(\ref{eq:4pt-1111-mij-dependence}) comes from the $U_0$-charge terms, see eq.~(\ref{ufactor}) and the comment below eq.~(\ref{eq:111-solution}); this fixes the $u_i$ dependence  of $C(u_1,u_2, x_4, \mathcal{S})$ to be 
\be 
C(u_1, u_2, x_4, \mathcal S) = \prod_{\alpha=1,2} u_\alpha \, (u_\alpha-1) \, (u_\alpha-x_4) \,  C(x_4, \mathcal S) \ . 
\ee

It is also straight-forward to confirm that the correlator satisfies indeed the incidence relation 
\eqref{eq:magic}. For our choice of insertion points this amounts to 
\begin{align}
& \Bigl\langle (\xi^-(z) + z \,  \xi^+(z)) \, \prod_{\alpha=1}^{2}W(u_\alpha) V_{m_1^1,m_2^1}^{1}(0;0) V_{m_1^2,m_2^2}^{1}(1;1) V_{m_1^3,m_2^3}^{1}(\infty;\infty) V_{m_1^4,m_2^4}^{1}(x_4;z_4) \Bigr\rangle \nonumber \\
& \qquad = \sum_{i=1}^4 \bigl\langle m_2^i+\tfrac{1}{2} \bigr\rangle = 0 \ , 
\end{align}
where we have used eq.~\eqref{eq:4pt-1111-system-solved}. 
\smallskip

This analysis takes care of the constraints that come from considering correlators with the insertion of $\xi^\pm$. We can also determine the constraints that come from the insertion of $\eta^\pm$ into the correlators, following the general method of Section~\ref{etaconstraints}; the analysis is a bit technical though, and we have therefore relegated it to Appendix~\ref{app:eta1111}. The most interesting novelty relative to the situation for the three-point functions of Section~\ref{sec:3pt-111} is that we cannot directly eliminate the unknown correlators $A^\pm_\alpha$, and that they actually turn out to involve also derivatives of delta-functions, not just delta-functions, see eq.~(\ref{eq:unknowns-ansatz}). As a side-product we also deduce the constraint that $\mathcal{S}$ has to equal $\mathcal{S}=0$, see eq.~(\ref{Scond}), which is just the $U_0$-charge condition. Note that $\mathcal{S}=0$ is in particular compatible with two of the $j_i = m_1^i - m_2^i$ to be equal to $j_i=\frac{1}{2}$, and the other two to be equal to $j_i = -\frac{1}{2}$.

\section{Conclusions}\label{sec:conclusions}

In this paper we have analysed the world-sheet correlation functions of string theory on ${\rm AdS}_3 \times {\rm S}^3 \times \mathbb{T}^4$ with minimal NS-NS flux ($k=1$), and made the matching with those of the dual symmetric orbifold theory manifest. The main advances relative to \cite{Eberhardt:2019ywk} are that:
\begin{list}{(\roman{enumi})}{\usecounter{enumi}}
\item Our analysis was  done directly in the hybrid formalism, using the free field realisation of $\mathfrak{psu}(1,1|2)_1$ in terms of symplectic bosons and fermions. (The analysis of \cite{Eberhardt:2019ywk} was done in the NS-R formalism, and only made use of the bosonic $\mathfrak{sl}(2,\mathds{R})_{k+2}$ symmetry.)
\item As a consequence we could derive stronger results: in particular, we could show that the solution is delta-function localised, see (\ref{corrstructfinal}) and \eqref{general-correlator}. (In \cite{Eberhardt:2019ywk} it was only shown that such a delta-function localised solution exists, but it was not clear (and probably also not true) that this is the only solution of the $\mathfrak{sl}(2,\mathds{R})_{k+2}$ Ward identities.)
\item These special properties of the correlation functions have a nice geometric origin, namely the `incidence relation' (\ref{eq:magic}), that suggests a close relation to a twistor string theory description, see Section~\ref{twstr}.
\end{list}
There are a number of open problems that remain. In particular, our analysis only concerns the chiral correlation functions, and we therefore cannot fix directly the full world-sheet amplitudes. However, given the results of this paper, it should now be possible to solve the bootstrap equations for the $\mathfrak{psu}(1,1|2)_1$ WZW model, and thereby fix the undetermined constants, say in (\ref{full-corr}). Once this is done, one should also add in the ghost (and fermion) contributions, and confirm that one obtains indeed the correct symmetric orbifold correlators (including normalisation, etc.). 

In this paper we have only considered the situation on the world-sheet sphere. However, most of the arguments also go through for higher genus; this will appear elsewhere \cite{Bob}. We have also only analysed the case of ${\rm AdS}_3 \times {\rm S}^3 \times \mathbb{T}^4$; it should be possible to do a similar analysis for ${\rm AdS}_3 \times {\rm S}^3 \times {\rm S}^3 \times {\rm S}^1$, for which the relevant WZW model is associated to $\mathfrak{d}(2,1|\alpha)_k$. The tensionless limit for that theory arises if the level of one of the $\mathfrak{su}(2)$ factors equals one, say $k^+=1$, and the world-sheet theory has a free field realisation in terms of symplectic bosons and fermions if also the other $\mathfrak{su}(2)$ level equals one, $k^+=k^-=1$, i.e.\ $\alpha=1$ and $k=\frac{1}{2}$ \cite{Eberhardt:2019niq}. Another class of backgrounds that should be accessible by these methods are the quotients of ${\rm AdS}_3 \times {\rm S}^3 \times \mathbb{T}^4$ that were recently studied in \cite{Eberhardt:2020bgq}.

The free field description of the world-sheet theory which we have developed and utilised in this paper should also allow one to attack other problems. For example, it should not be too difficult to study the D-branes of this world-sheet theory in terms of the free field description, see e.g.\ \cite{Kawai:2001ur,Gaberdiel:2006pp,Quella:2007sg}. This should give access to some non-perturbative aspects of the dual symmetric orbifold theory, and thereby shed light on the extent to which this duality holds beyond perturbation theory. 

Another direction which would be physically very important is to use the free field description to construct the vertex operator which adds an infinitesimal amount of RR 3-form flux (in the $g_s\rightarrow 0$ limit). It would be interesting to understand how to treat such a perturbation about the tensionless limit and, in particular, whether this modifies the delta function localisation of the correlators.

\acknowledgments

We thank Lorenz Eberhardt for useful conversations and comments on a draft of this paper. The work of AD, MRG and BK was supported by the Swiss National Science Foundation through a personal grant and via the NCCR SwissMAP. The work of RG is supported in part by the J. C. Bose Fellowship of the DST-SERB as well as in large measure by the framework of support for the basic sciences by the people of India.

\appendix 

\section{Spectrally flowed representations}\label{app:spectral-flow}

In this appendix we shall describe various spectrally flowed representations in more detail. We begin with the $\hat{\sigma}^2$-spectrally flowed vacuum representation. 

\subsection{Another vacuum representation}\label{app:W}

We want to show that the state defined by (\ref{Wdef}) is indeed the vacuum state with 
respect to $\mathfrak{psu}(1,1|2)$. First we note that the $\hat{\sigma}$ spectral flow acts 
 on the free fields as 
\begin{align}
\hat{\sigma} (\eta^\pm_r) = \eta^\pm_{r-1/2} \ , \qquad & \hat{\sigma} (\xi^\pm_r) = \xi^\pm_{r+1/2}  \\
\hat{\sigma} (\chi^\pm_r) = \chi^\pm_{r+1/2} \ , \qquad & \hat{\sigma} (\psi^\pm_r) = \psi^\pm_{r-1/2} \ . 
\end{align}
The action on the $\mathfrak{psu}(1,1|2)_1$ and $\mathfrak{u}(1)$ generators is then
\begin{subequations}
\begin{align}
\hat \sigma(J^3_m) & = J^3_m  \ , & \hat \sigma(J^\pm_m) &= J^{\pm}_{m} \ , \label{Jhat} \\
\hat \sigma(K^3_m) & = K^3_m \ , & \hat \sigma(K^\pm_m) &= K^{\pm}_{m} \ , \label{Khat} \\
\hat \sigma(S^{\alpha \beta \pm}_m) & = S^{\alpha \beta \pm}_{m \pm 1} \ , & \hat \sigma(U_m) &= U_m + \tfrac{1}{2}\delta_{m,0} \ , \label{Uhat}\\
\hat \sigma(V_m) & = V_m + \tfrac{1}{2} \delta_{m,0} \ , & \hat \sigma(Z_m) & = Z_m + \delta_{m,0} \ . \label{Vhat}
\end{align}
\end{subequations}
It therefore (almost) leaves the $\mathfrak{psu}(1,1|2)$ generators invariant, while it shifts the eigenvalues of $U_0$ and $V_0$. 

In order to see that (\ref{Wdef}) is indeed the vacuum state we first note that because of (\ref{Jhat}) and (\ref{Khat}) this is the case with respect to the bosonic subalgebra $\mathfrak{sl}(2,\mathbb{R})_1\oplus\mathfrak{su}(2)_1$; here we have used that the fermionic excitations do not modify this property, as one can easily see from the free field realisation. Note that the fermionic generators are required to guarantee that the state is also annihilated by the $S^{\alpha\beta\gamma}_n$ modes with $n\geq 0$; in particular, since $\hat \sigma(S^{\alpha \beta -}_m) = S^{\alpha \beta -}_{m-1}$, this requires that 
\be
S^{\alpha \beta -}_{m-1} \, \psi^+_{-3/2} \, \psi^-_{-3/2} \,  \psi^+_{-1/2} \,  \psi^-_{-1/2} |0\rangle  = 0 \ , \qquad m\geq 0  \ ,
\ee
as one verifies again directly from the free field realisation. The other supercharge generators work similarly. Finally, the $U_0$ charge of $|0\rangle^{(1)}$ follows directly from the spectral flow property (\ref{Uhat}), while for $V_0$ we use that 
\be
V_0 \, \psi^+_{-3/2} \, \psi^-_{-3/2} \,  \psi^+_{-1/2} \,  \psi^-_{-1/2} |0\rangle  = - 2 \, \psi^+_{-3/2} \, \psi^-_{-3/2} \,  \psi^+_{-1/2} \,  \psi^-_{-1/2} |0\rangle \ , 
\ee
since each $\psi^\alpha$ generator carries $V_0$ charge $-1/2$. Together with (\ref{Vhat}) this then shows that $V_0 |0\rangle^{(1)} = - |0\rangle^{(1)}$, and hence leads to (\ref{Wcharges}).

\subsection{Twisted sector ground states}\label{app:twisted}

In this appendix we identify the twisted sector ground state of the dual CFT in terms of our hybrid description. Let us start by recalling that under the usual spectral flow the operators transform as 
\begin{subequations}
\begin{align}
\sigma^w (J^3_m)&=J^3_m+\tfrac{kw}{2}\delta_{m,0}\ , \label{eq:spectral flow a}\\
\sigma^w(J^\pm_m)&=J^\pm_{m\mp w}\ , \label{eq:spectral flow b}\\
\sigma^w(K^3_m)&=K^3_m+\tfrac{k w}{2}\delta_{m,0}\ , \label{eq:spectral flow c}\\
\sigma^w(K^\pm_m)&=K^\pm_{m \pm w}\ , \label{eq:spectral flow d}\\
\sigma^w(S^{\alpha\beta\gamma}_m)&=S^{\alpha\beta\gamma}_{m+\frac{1}{2}w(\beta-\alpha)}\ .\label{eq:spectral flow e}
\end{align}
\end{subequations}
In addition, the energy-momentum tensor transforms as 
\be 
\sigma^w(L_m)= L_m+w(K_m^3-J_m^3)\ . \label{eq:spectral flow f} 
\ee
 In the hybrid formalism, the mass-shell condition is 
\be
L_0=0 \ .
\ee
We also note that on the highest weight states of the above (unflowed) representation $L_0=0$. This follows from the fact that the Casimir of $\mathfrak{psu}(1,1|2)$ vanishes on the short representation, see eq.~(3.25) of  \cite{Eberhardt:2018ouy}; it also follows from the observation that the $4$ free fermions contribute $\Delta_{\rm f} = \frac{4}{16}=\frac{1}{4}$ to the ground state energy --- they are in the R sector --- while the $4$ symplectic bosons contribute $\Delta_{\rm b} = - \frac{4}{16}= - \frac{1}{4}$, see e.g.\ Section~4.2 of \cite{Gaberdiel:2018rqv} so that the total ground state energy is zero. Thus the mass-shell condition in the $w$-spectrally flowed sector is 
\be\label{massshell}
N + w (n-m) = 0 \ , 
\ee
where $N$ is the excitation number before spectral flow, while $m$ and $n$ are the $J^3_0$ and $K^3_0$ eigenvalues before spectral flow. 

\subsubsection[The case of odd \texorpdfstring{$w$}{w}]{\boldmath The case of odd \texorpdfstring{$w$}{w}} 

For odd spectral flow $w$, the twisted sector ground state should have conformal dimension \be\label{hodd}
h = \frac{w^2-1}{4w} \ , 
\ee
and define a singlet (highest weight state) with respect to the $\mathfrak{su}(2)_1$ algebra. From the world-sheet perspective the relevant state is 
\be\label{twistedodd}
[\Phi_w]^{\sigma^w} \ , \qquad \hbox{with} \qquad 
\Phi_w = \chi^-_{-\frac{w-1}{2}} \psi^-_{-\frac{w-1}{2}}  \cdots \chi^-_{-1} \psi^-_{-1} \, \chi^-_{0} \psi^-_{0} |m_1,m_2\rangle \ , 
\ee
where $m_1,m_2$ are fixed by the mass-shell condition and the requirement that $Z_0=0$, 
\be
m_1 + m_2  = - \frac{w^2+1}{4w} \ ,  \qquad 
m_1 - m_2 = \frac{1}{2} \ . 
\ee
Indeed, since the $K^3_0$ eigenvalue before spectral flow is $n = - \frac{w}{2}$, the $K^3_0$ eigenvalue after spectral flow is zero. (In fact, the state transforms as the vacuum with respect to $\mathfrak{su}(2)_1$.) Similarly, the $J^3_0$ eigenvalue before spectral flow equals $m= m_1 + m_2 = - \frac{w^2+1}{4w}$, and thus (\ref{massshell}) is satisfied with excitation number $N = \frac{(w-1)(w+1)}{4}$.  The state is also physical since it is annihilated by $L_0$ and ${\cal Q}_0$, given that the modes ${\cal Q}_n$ are explicitly 
\be
{\cal Q}_n = 2 \sum_{r,s,p,q} (r-s) \, \xi^+_r \xi^-_s \chi^+_p \chi^-_q \, \delta_{r+s+p+q,n} \ . 
\ee
On the $\sigma^w$ twisted sector state (\ref{twistedodd}) the fermionic modes are only non-zero provided that $p+q\leq -1$, and as a consequence $r+s\geq n+1$. In particular, ${\cal Q}_0$ therefore annihilates the state, while for ${\cal Q}_{-1}$ the contribution with $r=\frac{w}{2}$ and $s=-\frac{w}{2}$ survives, leading to (\ref{Qm1odd}). 

 \subsubsection[The case of even \texorpdfstring{$w$}{w}]{\boldmath The case of even \texorpdfstring{$w$}{w}} 
 
 For even spectral flow, the analogue of (\ref{hodd}) is 
\be
h = \frac{w}{4} \ , 
\ee
and the ground state of the twisted sector is two-fold degenerate, with the two states transforming as a doublet with respect to $\mathfrak{su}(2)$. The two states are then given by $[\Phi^\pm_w]^{\sigma^w}$, with 
\begin{eqnarray}
\Phi^+_{w}  & = &  \chi^-_{-\frac{w}{2}+1} \psi^-_{-\frac{w}{2}+1}  \cdots \chi^-_{-1} \psi^-_{-1} \, \chi^-_{0} \psi^-_{0} \, |m_1,m_2\rangle \label{twistedevenp}\\ 
\Phi^-_{w}  & = &  \chi^-_{-\frac{w}{2}} \psi^-_{-\frac{w}{2}} \chi^-_{-\frac{w}{2}+1} \psi^-_{-\frac{w}{2}+1}  \cdots \chi^-_{-1} \psi^-_{-1} \, \chi^-_{0} {\psi}^-_{0} \, |m_1,m_2\rangle \ . \label{twistedevenm}
\end{eqnarray}
In this case the mass-shell condition and the requirement that $Z_0=0$ are 
\be
m_1 + m_2 = - \frac{w}{4} \ ,  \qquad 
m_1 - m_2 = \frac{1}{2} \ . 
\ee

\section{On Covering Maps}\label{app:cover}

In this appendix we explain how the conditions on the covering map determine the polynomials $p^\pm(z)$ up to two overall scale factors as well as two coefficients. Since we can factor out the two overall scale factors, this, in particular, implies that the conditions on the covering map can be written in terms of ratios of expansion coefficients, see the discussion below eq.~(\ref{eq:ab-eqns}). Let us start by denoting the coefficients of $p^\pm(z)$ as 
\begin{equation}
p^{-}(z) =\sum_{k=0}^N a_kz^k \ , \qquad \qquad  p^{+}(z) =\sum_{k=0}^N b_kz^k \ .
\end{equation}
Then the constraint \eqref{eq:ppm-constraint} becomes 
\begin{equation}
\sum_{k,l=0}^N (k-l)a_kb_lz^{k+l-1} = \sum_{j=1}^{2N-1}C_{j-1}z^{j-1} \, .
\label{eq:ab-constr}
\end{equation}
Here the right hand side is the expansion of the polynomial $C\prod_{i=1}^n(z-z_i)^{w_i-1}$, with 
$C_{2N-2}=C$ being the only unknown. The left hand side has the $(2N+2)$ unknowns $\{a_j, b_j\}$.

We can eliminate $C$ by considering the $(2N-2)$ equations that come from equating powers of $z$ on both sides of  \eqref{eq:ab-constr},
\begin{equation} 
\sum_{l=0}^{{\rm min}{(j,N)}} (j-2l)\, a_{j-l}b_l = \tilde{C}_{j-1}(a_Nb_{N-1}-a_{N-1}b_N)  \ ,
\label{eq:ab-eqns}
\end{equation}
with $\tilde{C}_{j}(\{z_i, w_i\})=\frac{C_j}{C}$ and $j=1,\ldots, (2N-1)$.  
Since \eqref{eq:ab-eqns} is homogeneous in the $\{a_j, b_j\}$, we can take out a common factor of, say, $(a_Nb_N)$, and write these equations purely in terms of the $2N$ variables $\{\alpha_j=\frac{a_j}{a_N}, \beta_j=\frac{b_j}{b_N} \}$ for $j=0,\ldots, N-1$. 
We can fix two of these, say $(\alpha_0, \beta_0)$, and parametrically solve these $(2N-2)$ equations for the remaining $(2N-2)$ variables. This algebraic system of equations will generically admit a finite number of discrete solutions for the $\{\alpha_j, \beta_j\}$ for any fixed $(\alpha_0, \beta_0)$. 

We have thus obtained a family of covering maps which have the right branching behaviour at $z=z_i$ and are expressible in terms of the two parameters $(\alpha_0, \beta_0)$, together with the individual scaling factors of $(a_N, b_N)$. We can thus write the solutions to  \eqref{eq:ab-constr} as
\begin{equation}
p^{-}(z) =a_N\, q^-(\alpha_0, \beta_0; z) \ , \qquad \qquad  p^{+}(z) =b_N \, q^+(\alpha_0, \beta_0; z)\ .
\label{eq:ppm-sol}
\end{equation}
Thus we have a solution for the covering map in terms of a common scale factor and three unknowns 
$(\frac{a_N}{b_N}, \alpha_0, \beta_0)$. 

We can fix these three unknowns by demanding that 
\begin{equation}
p^-(z_i)+x_i\,p^+(z_i) = 0
\label{eq:gamma-constr-sp}
\end{equation}  
for, say, $i=1,2,3$. Again, as algebraic equations we generically will have finitely 
many solutions. The covering map is thus determined up to these discrete choices. In particular,
$\Gamma(z_i)$ for $(i=4,\ldots, n)$ are now fixed to discretely many values.  

\section{Using the incidence relation to determine correlators explicitly}\label{app:explicit}

The insights of Section~\ref{sec:twistor} do not just exhibit the underlying structure of the correlation functions. They also provide a quite powerful method for the explicit calculation of the correlation functions. In this appendix we demonstrate this for the example of a four-point function with $w_i=2$ for all $i$.

\subsection{A simple example}

Let us set without loss of generality $x_1=z_1=0$, $x_2=z_2=1$, and $x_3=z_3=\infty$. Then the covering map with branching $w_i=2$ for all $i$ is given by
\begin{equation}
\Gamma(z)=\frac{z^2\left(\pm z\sqrt{z_4^2-z_4+1}-z\,z_4-z+3z_4\right)}{\pm(3z-2)\sqrt{z_4^2-z_4+1}+3z\,z_4-z_4+2} \ , 
\end{equation}
with the two signs corresponding to the two possible covering maps. Moreover, this covering map only sends $z_4$ to $x_4$ provided that
\begin{equation}
x_4=\Gamma(z_4)=z_4\left(\pm\,2(z_4-1)\sqrt{z_4^2-z_4+1}+2z_4^2-3z_4+2\right)\ .
\label{2222-requirement}
\end{equation}
We wish to show that the correlators are only nonvanishing on this locus. The degree of the covering map is $N=3$, and so $P^+$ and $P^-$ are order 3 polynomials. Let us write
\begin{equation}
P^{\pm}(z)=\sum_{m=0}^{3}c^{\pm}_mz^m\ .
\end{equation}
The constraints (\ref{polynomial-constraint}) imply that 
\begin{equation}
\begin{split}
P^-(z_i)+x_iP^+(z_i)&=\sum_{m=0}^{3}\bigl(c^-_m+x_ic^+_m\bigr)\, z_i^m=0\ ,\\
\partial\bigl(P^-(z)+x_iP^+(z)\bigr)\Big|_{z=z_i}&=\sum_{m=0}^{3}m\, 
\bigl(c^-_m+x_ic^+_m\bigr)\, z_i^{m-1}=0\ , \end{split}
\end{equation}
for $i=1,2,3,4$. Since we have set $x_1=z_1=0$, $x_2=z_2=1$, and $x_3=z_3=\infty$,\footnote{As always, taking insertion points to infinity requires some care. In this case, one must take the limit in such a way that $x_3/z_3^2$ remains constant.} we obtain $c^-_0=c^-_1=0$ and $c^+_2=c^+_3=0$, while the remaining system is given by
\begin{equation}
\begin{split}
c^-_2+c^-_3+c^+_0+c_1^+&=0\\
2c_2^-+3c_3^-+c_1^+&=0\\
c^-_2z_4^2+c^-_3z_4^3+c^+_0x_4+c_1^+x_4z_4&=0\\
2c_2^-z_4+3c_3^-z_4^2+c_1^+x_4&=0\ .
\end{split}
\end{equation}
This is a homogeneous system of four equations and four unknowns. It only has a nontrivial solution if its determinant vanishes, namely if $x_4^2-4x_4z_4+6x_4z_4^2-4x_4z_4^3+z_4^4=0$. In fact, one can massage this system to show that
\begin{equation}
\bigl(x_4^2-4x_4z_4+6x_4z_4^2-4x_4z_4^3+z_4^4\bigr)\, c_m^{\pm}=0\ .
\end{equation}
The polynomial prefactor vanishes precisely when \eqref{2222-requirement} is satisfied, and thus the coefficients $c_m^{\pm}$ are localised on the locus $x_4=\Gamma(z_4)$,
\begin{equation}
c_m^{\pm}\propto\sum_{\Gamma}(c_m^{\pm})_\Gamma\,\, \delta(x_4-\Gamma(z_4))\ .
\end{equation}
Since the correlators are simply linear combinations of the coefficients of $P^{\pm}(z)$ (see \eqref{polynomial-to-correlator}), we conclude that the correlators localise on the same locus.

\subsection[Delta-function localisation for four-point functions]{\boldmath Delta-function localisation for four-point functions}\label{app:4ptdelta}

In this appendix we give a direct argument to show that the four-point functions are delta-function localised.  Let $\vec{v}$ be the $2N+2$-dimensional column vector containing the coefficients of $P^{\pm}(z)$, and let $M$ be the $(2N+2)\times(2N+n-2)$ matrix implementing the linear system of constraints, see eq.~(\ref{polynomial-constraint2})
\begin{equation}
\partial^{\ell-1}\bigl (P^-(z)+x_iP^+(z)\bigr)\Bigl|_{z=z_i}=0\ ,\qquad \ell=1,\ldots,w_i\ ,
\label{polynomial-linear-system}
\end{equation}
which we may write as a matrix equation, $M\vec{v}=0$. (Recall that $\sum_i w_i = 2 N + n-2$.)
For $n=4$, $M$ is a square matrix, and let $A$ be the adjugate matrix of $M$, i.e.\ the matrix such that $AM=\det{M}\,\textbf{1}$. Then we can multiply both sides of our system by $A$ --- note that the entries of both $M$ and $A$ are regular functions of $x_i$ and $z_i$ --- and we obtain
\begin{equation}
\left(\det{M}\right)\,\vec{v}=0\implies\bigl(\det{M}\bigr)\, P^{\pm}(z)=0\ .
\end{equation}
Using eq.~(\ref{polynomial-to-correlator}) it then also follows that 
\be
\left(\det{M}\right)\,\Bigl\langle \prod_{\alpha=1}^{2}W(u_{\alpha})\prod_{i=1}^{4}V_{m_1^i,m_2^i}^{w_i}(x_i;z_i) \Bigr\rangle=0\ .
\ee
Now, let us assume that we have fixed $x_i$ for $i=1,2,3$ and $z_i$ for $i=1,2,3,4$, so that $x_4$ is a free parameter. Since $\det{M}$ is a polynomial in $x_4$, we can write
\begin{equation}\label{detM}
\det{M}=C\, \prod_{\Gamma} \bigl(x_4-\Gamma(z_4)\bigr) \ ,
\end{equation}
where the product is over the different covering maps, and $C$ is a non-zero constant. Here we have used that $\det{M}=0$ is necessary in order for the covering map to exist --- recall that (\ref{polynomial-linear-system}) are also the constraints that characterise the covering map.\footnote{We have also checked (\ref{detM}) explicitly, and at least for generic choices of $x_i$ and $z_i$ this identity is true. For special cases the determinant $\det(M)$ may actually be a higher order polynomial in $x_4$, but then the additional factors that appear on the right hand side involve $x_4$ or $(x_4-1)$, which describe various degeneration limits.}
Putting these two identities together we conclude that  
\be
\prod_{\Gamma} \bigl(x_4-\Gamma(z_4)\bigr) \, \, \Bigl\langle \prod_{\alpha=1}^{2}W(u_{\alpha})\prod_{i=1}^{4}V_{m_1^i,m_2^i}^{w_i}(x_i;z_i) \Bigr\rangle=0\ ,
\ee
and the correlator must therefore be a sum of delta-functions, 
\begin{equation}
\Bigl\langle \prod_{\alpha=1}^{2}W(u_{\alpha})\prod_{i=1}^{4}V_{m_1^i,m_2^i}^{w_i}(x_i;z_i) \Bigr\rangle=\sum_{\Gamma}C_{\Gamma}\,\delta\left(x_4-\Gamma(z_4)\right)\ ,
\end{equation}
where $C_{\Gamma}$ contains again the dependence on the other variables.

\section{\boldmath The \texorpdfstring{$\eta^\pm$}{eta+-} analysis for the four-point function with \texorpdfstring{$w_i=1$}{wi = 1}}\label{app:eta1111}

In this appendix we analyse the constraints that come from the insertion of the $\eta^\pm$ fields into the correlator for the case of the four-point function with $w_i=1$, following the general method outlined in Section~\ref{etaconstraints}. In particular, eq.~(\ref{etareg}) leads to the recursion relations 
\be 
2 \, m_2^i \,  \bigl\langle m_2^i - \tfrac{1}{2} \bigr\rangle + 2 \sum_{j \neq i} \frac{x_i - x_j}{z_i - z_j} \, m_1^j \bigl\langle m_i^j + \tfrac{1}{2} \bigr\rangle + \sum_{\alpha =  1,2} \frac{A^-_\alpha + x_i A^+_\alpha}{z_i-u_\alpha} = 0 \ . 
\label{eq:4pt-1111-eta-recursions}
\ee
Note that unlike what happened for the three-point functions in Section \ref{sec:3pt-111}, this time we cannot simply eliminate the unknowns $A^\pm_{\alpha}$ and obtain a relation involving only correlators of the form \eqref{eq:4pt-1111}. In fact, \eqref{eq:4pt-1111-eta-recursions} amounts to four equations in the four unknowns $A^\pm_{\alpha}$.

One strategy to proceed is to derive relations for the correlators $A^\pm_{\alpha}$ by inserting an additional $\xi^\pm$ field into the $A^\pm_{\alpha}$ correlator, and using the techniques of Section~\ref{sec:Ward}. To be specific, for the correlator $A^+_1$, we consider 
\begin{subequations}
\begin{align}
\braket{ \xi^-(z) (\eta^+_{\frac{1}{2}}W)(u_1)W(u_2) VVVV} &= - \sum_{i=1}^4 \frac{x_i}{z-z_i}A^+_1(m_2^i + \tfrac{1}{2})  \ , \\
\braket{ \xi^+(z) (\eta^+_{\frac{1}{2}}W)(u_1)W(u_2) VVVV} &= \sum_{i=1}^4 \frac{A^+_1(m_2^i + \tfrac{1}{2})}{z-z_i} \ , 
\end{align}
\end{subequations}
where we have adopted the notation
\be 
A^+_1(m_2^1 + \tfrac{1}{2}) = \Bigl\langle (\eta^+_{\frac{1}{2}}W)(u_1) \, W(u_2) \, V_{m_1^1,m_2^1+\frac{1}{2}}^{1}(x_1;z_1) \, \prod_{i=2}^{4}V_{m_1^i,m_2^i}^{1}(x_i;z_i)  \Bigr\rangle
\ee
and similarly for $A^+_1(m_2^i + \tfrac{1}{2})$ with $i=2,3,4$. 
Since
\be 
\xi^-(z) \bigl(\eta^+_{\frac{1}{2}}W\bigr)(u_\alpha) = \bigl(\xi^-_{-\frac{1}{2}} \eta^+_{\frac{1}{2}}W\bigr)(u_\alpha) + \mathcal{O}(z-u_\alpha) = - W(u_\alpha) + \mathcal{O}(z-u_\alpha) \ , 
\ee
imposing regularity at $u_1$ and $u_2$ leads to 
\begin{subequations}
\begin{align}
& \sum_{i=1}^4 \frac{x_i \, A_1^+(m_2^i + \tfrac{1}{2})}{u_1-z_i}  = \braket{ \ \dots \ } \ , & & \sum_{i=1}^4 \frac{x_i \, A_1^+(m_2^i + \tfrac{1}{2})}{u_2-z_i} = 0 \\
& \sum_{i=1}^4 \frac{A_1^+(m_2^i + \tfrac{1}{2})}{u_1-z_i} = 0 \ , & & \sum_{i=1}^4 \frac{A_1^+(m_2^i + \tfrac{1}{2})}{u_2-z_i} = 0 \ , 
\end{align}
\end{subequations}
where 
\be 
\braket{ \ \dots \ } := \Bigl\langle \prod_{\alpha=1}^{2}W(u_\alpha) \prod_{i=1}^{4}V_{m_1^i,m_2^i}^{1}(x_i;z_i) \Bigr\rangle \ . 
\ee
Similar relations can be derived for $A^-_1$, $A^+_2$ and $A^-_2$, and the resulting relations take the form  
\begin{subequations} 
\begin{align}
& \sum_i \frac{x_i \, A^+_\alpha(m_2^i+\tfrac{1}{2})}{u_\beta - z_i}= \braket{ \ \dots \ } \, \delta_{\alpha \beta}  \ , & & \sum_i \frac{A^-_\alpha(m_2^i+\tfrac{1}{2})}{u_\beta - z_i}= \braket{ \ \dots \ } \, \delta_{\alpha \beta}  \ , \\
& \sum_i \frac{A^+_\alpha(m_2^i+\tfrac{1}{2})}{u_\beta  - z_i}=0 \ , & & \sum_i \frac{x_i \, A^-_\alpha(m_2^i+\tfrac{1}{2})}{u_\beta - z_i}=0 \ ,  
\end{align}
\label{eq:unknowns-eqs-summary}
\end{subequations}
where $\alpha, \beta \in \{1,2 \}$. 

Next we solve the combined system of eqs.~(\ref{eq:4pt-1111-eta-recursions}) and (\ref{eq:unknowns-eqs-summary}). Performing the shift $m_2^i \mapsto m_2^i + \tfrac{1}{2}$ 
in (\ref{eq:4pt-1111-eta-recursions}), and summing over $i = 1, \dots, 4$, we deduce 
\be
\sum_{i=1}^4 m_2^i \, \braket{ \ \dots \ }  +  \sum_{i=1}^4 \sum_{j \neq i} \frac{x_i - x_j}{z_i - z_j} \, m_1^i \bigl\langle m_1^i + \tfrac{1}{2} ; m_2^j+\tfrac{1}{2} \bigr\rangle = 0 \ ,
\label{eq:4pt-1111-rec-simple}
\ee
where we have used (\ref{eq:unknowns-eqs-summary}). Note that we have exchanged the roles of $i$ and $j$ in the second term of \eqref{eq:4pt-1111-rec-simple}. 
We can further simplify the result by realising that \eqref{eq:4pt-1111-zi-regularity} implies 
\be 
- \braket{ \ \dots \ }  = \sum_{j\neq i} \frac{x_i-x_j}{z_i-z_j} \bigl\langle m_1^i + \tfrac{1}{2}; m_2^j + \tfrac{1}{2} \bigr\rangle \ , \qquad i = 1, \dots, 4 \ . 
\ee 
Thus eq.~\eqref{eq:4pt-1111-rec-simple} becomes 
\be 
\sum_{i=1}^4 (m_1^i-m_2^1) \braket{ \ \dots \ } = \mathcal S  \braket{ \ \dots \ } = 0  \ , 
\ee
where we have used eq.~(\ref{eq:4pt-1111-mij-dependence}). Thus we can deduce that 
\be \label{Scond}
C(u,\mathcal{S}) = C(u) \,  \delta_{\mathcal S,0} \ , \qquad \text{or equivalently} \qquad \braket{ \ \dots \ } \propto \delta_{\mathcal S, 0} \ ,
\ee
which is precisely the $U_0$ charge conservation constraint discussed in Section \ref{sec:U0charge}. 

\subsection{A consistent solution}

We have seen in Section~\ref{sec:twistorrel} that the correlators themselves are delta-function localised. One may then ask what the localisation property of the $A^\pm_\alpha$ correlators are, and whether these various relations have in fact a consistent solution. In order to study the functional dependence of these correlators it is useful to take advantage of the fact that 
\be
J_0^+ V^1_{m_1, m_2}(x,z) = \partial_x V^1_{m_1, m_2}(x,z) \ . 
\ee
In terms of the free fields $\xi^+$ and $\eta^+$, see eq.~\eqref{eq:u112-algebra}, this becomes 
\be 
J_0^+ \bigl[| m_1^i, m_2^i \rangle \bigr]^{\sigma^1} = \eta^+_{-\frac{1}{2}} \bigl[ | m_1^i,m_2^i+\tfrac{1}{2}\rangle \bigr]^{\sigma^1} + 2 \, m_1^i \,  \xi^+_{-\frac{1}{2}} \bigl[ | m_1^i+\tfrac{1}{2},m_{2}^i \rangle \bigr]^{\sigma^1} \ . 
\label{eq:J0p-on-ket}
\ee
Inserting \eqref{eq:J0p-on-ket} into a correlation function,\footnote{This procedure is somewhat analogous to the way in which the Knizhnik-Zamolodchikov equation \cite{Knizhnik:1984nr} is derived: we write the derivative operator $\partial_x = J^+_0$ in terms of bilinears of (in our case free) fields, and then use contour deformation techniques to rewrite this bilinear expression of the free fields. Note though that for the actual Knizhnik-Zamolodchikov equation the derivative operator is a bilinear of currents (via the Sugawara construction), while here they are the free symplectic boson fields.} we obtain the differential equation  
\begin{multline} 
\partial_{x_i} \braket{ \ \dots \ } = \Bigl\langle \prod_{\alpha=1}^{2}W(u_\alpha) \Bigl(\eta^+_{-\frac{1}{2}}V^1_{m_1^i, m_2^i+\frac{1}{2}}\Bigr)(x_i,z_i) \prod_{j\neq i} V^1_{m_1^j, m_2^j}(x_j, z_j)\Bigr\rangle \\
+ 2 \, m_1^i \,  \Bigl\langle \prod_{\alpha=1}^{2}W(u_\alpha) \Bigl(\xi^+_{-\frac{1}{2}}V^1_{m_1^i+\frac{1}{2}, m_2^i}\Bigr)(x_i,z_i) \prod_{j\neq i} V^1_{m_1^j, m_2^j}(x_j, z_j)\Bigr\rangle \ . 
\label{eq:KZ-implicit}
\end{multline}
The two terms on the right-hand-side can be simplified by rewriting them in terms of a contour integral, which we may then deform to encircle the other insertion points. This leads to 
\begin{align}
& \Bigl\langle \prod_{\alpha=1}^{2}W(u_\alpha) \Bigl(\xi^+_{-\frac{1}{2}}V^1_{m_1^i, m_2^i}\Bigr)(x_i,z_i) \prod_{j\neq i} V^1_{m_1^j, m_2^j}(x_j, z_j)\Bigr\rangle = - \sum_{j \neq i}\frac{\braket{m_2^j+\tfrac{1}{2}}}{(z_j-z_i)} \ , \\
& \Bigl\langle \prod_{\alpha=1}^{2}W(u_\alpha) \Bigl(\eta^+_{-\frac{1}{2}}V^1_{m_1^i, m_2^i}\Bigr)(x_i,z_i) \prod_{j\neq i} V^1_{m_1^j, m_2^j}(x_j, z_j)\Bigr\rangle \nonumber \\
& \hspace{180pt}  = -\sum_{\alpha=1,2} \frac{A^+_\alpha}{u_\alpha-z_i} - \sum_{j \neq i}\frac{2 m_1^j \braket{m_1^j+\tfrac{1}{2}}}{(z_j-z_i)} \ . 
\end{align}
Then eq.~\eqref{eq:KZ-implicit} becomes 
\begin{multline}
\partial_{x_i} \braket{\ \dots \ } + \sum_{\alpha=1,2}\frac{A^+_\alpha(m_2^i+\tfrac{1}{2})}{u_\alpha-z_i}  \\
+\sum_{j \neq i}\frac{2 m_1^j \braket{m_1^j+\tfrac{1}{2};m_2^i+\tfrac{1}{2}}}{(z_j-z_i)} +\sum_{j \neq i}\frac{2m_1^i \braket{m_1^i+\tfrac{1}{2};m_2^j+\tfrac{1}{2}}}{(z_j-z_i)} =0 \ . 
\label{eq:KZ-explicit}
\end{multline}
As a side-remark we note that multiplying eq.~\eqref{eq:KZ-explicit} by $x_i^{p+1}$ with $p\in \{-1,0,1\}$ and summing over $i=1, \dots, 4$, we recover the  global Ward identities of the spacetime CFT
\be 
\sum_{i=1}^4 \biggl[x_i^{p+1} \partial_{x_i} + (p+1)\left(m_1^i+m^i_2+\tfrac{1}{2}\right)\, x_i^p \biggr] \braket{\ \dots \ } = 0 \ . 
\ee
Here, however, we want to use eq.~\eqref{eq:KZ-explicit} to constrain the functional dependence of the different correlators. While we have not managed to show that there is a unique solution for the $A^\pm_\alpha$ correlators, we claim that there exists a solution with the ansatz 
\be
A^\pm_\alpha(m_2^i+\tfrac{1}{2}) = \mathcal{A}^\pm_{\alpha, i} \, \delta(z_4-x_4) + \mathcal{A}^{'\pm}_{\alpha,i} \,  \delta'(z_4-x_4) \ ,
\label{eq:unknowns-ansatz}
\ee
where $\mathcal{A}^\pm_{\alpha,i}$, and $\mathcal{A}^{'\pm}_{\alpha,i}$ are arbitrary functions of $u_1, u_2$ and $z_4$. The reason for this ansatz is that 
multiplying \eqref{eq:unknowns-eqs-summary} by $(z_4-x_4)$ one finds that 
\be 
(z_4-x_4) A^\pm_{\alpha}(m_2^i+\tfrac{1}{2}) 
\ee
has support only at $z_4 - x_4 = 0$, i.e.\ the determinant of the system vanishes for $z_4 = x_4$. Moreover, \eqref{eq:unknowns-ansatz} is motivated by the fact that the terms in the second line of \eqref{eq:KZ-explicit} are proportional to $\delta(z_4-x_4)$ while contributions proportional to $\delta'(z_4-x_4)$ are produced by taking a derivative of the correlator $\braket{\ \dots \ }$ in the first line. 

Making use of the distributional identity 
\be 
\delta'(x) f(x) = \delta'(x)f(0) - \delta(x)f'(0) \ ,
\ee
and writing the correlator $\braket{ \ \dots \ }$ in terms of an arbitray function $\mathcal F$ of $z_4, u_1$ and $u_2$, 
\be 
\braket{ \ \dots \ } = \mathcal F (z_4, u_1, u_2) \  \delta(z_4-x_4) \ , 
\ee
eqs.~\eqref{eq:unknowns-eqs-summary} become for $\alpha, \beta \in \{ 1 ,2 \} $
\begin{subequations}
\begin{align}
& \sum_i \frac{z_i \,  \mathcal{A}^{'\pm}_{\beta,i}}{u_\alpha - z_i} = 0 \ , & & \sum_i \frac{ \mathcal{A}^{'\pm}_{\beta,i}}{u_\alpha - z_i} = 0 \ , \\
& \sum_i \frac{z_i \, \mathcal{A}^{+}_{\beta,i}}{u_\alpha - z_i} - \frac{\mathcal{A}^{'+}_{\beta,4}}{u_\alpha-z_4} = \mathcal F (z_4, u_1, u_2) \, \delta_{\alpha \beta} \ , && \sum_i \frac{\mathcal{A}^+_{\beta,i}}{u_\alpha - z_i} = 0 \ , \\
&  \sum_i \frac{z_i \, \mathcal{A}^-_{\beta,i}}{u_\alpha - z_i} - \frac{\mathcal{A}^{'-}_{\beta,4}}{u_\alpha-z_4} = 0 \ , && \sum_i \frac{\mathcal{A}^-_{\beta,i}}{u_\alpha - z_i}= \mathcal F (z_4, u_1, u_2) \, \delta_{\alpha \beta}  \ ,
\end{align}
\label{eq:big-system-unknowns-1}
\end{subequations}
while eq.~\eqref{eq:4pt-1111-eta-recursions} (with $m_2^i \mapsto m_2^i + \tfrac{1}{2}$) implies 
\begin{subequations}
\begin{align}
& \sum_\alpha \frac{\mathcal{A}^{'-}_{\alpha,i} + x_i \,  \mathcal{A}^{'+}_{\alpha,i}}{z_i - u_\alpha} = 0 \ ,  \\
& \sum_{\alpha} \frac{\mathcal A^-_{\alpha,i} + z_i \, \mathcal A^+_{\alpha,i}}{z_i - u_{\alpha}} -\sum_\alpha  \frac{\mathcal A^{'+}_{\alpha,i}}{z_i - u_\alpha}\delta_{i,4} \nonumber \\
& \hspace{80pt} +2 (m_2^i+\tfrac{1}{2}) \braket{ \ \dots \ } + 2 \sum_{j\neq i} \frac{x_i-x_j}{z_i-z_j} m_1^j \braket{m_1^j + \tfrac{1}{2}; m_2^i + \tfrac{1}{2}} = 0 \ ,
\end{align}
\label{eq:big-system-unknowns-2}
\end{subequations}
where $i=1,2,3$. Finally, for the differential equation \eqref{eq:KZ-explicit} with $i=4$ to be obeyed, it is necessary that
\begin{subequations} 
\begin{align}
& \sum_\alpha \frac{\mathcal A^{'+}_{\alpha,4}}{u_\alpha - z_4} = - \mathcal F (z_4, u_1, u_2) \ ,  \\
& \sum_{\alpha} \frac{\mathcal A^+_{\alpha,4}}{u_\alpha-z_4} + \sum_{j \neq 4} \frac{2 m_1^j \braket{m_1^j+\tfrac{1}{2};m_2^4+\tfrac{1}{2}}}{z_j-z_4} + \sum_{j \neq 4} \frac{2 m_1^4 \braket{m_1^4+\tfrac{1}{2};m_2^j+\tfrac{1}{2}}}{z_j-z_4} = 0 \ . 
\end{align} 
\label{eq:big-system-unknowns-3}
\end{subequations}
\noindent Eqs.~\eqref{eq:big-system-unknowns-1},  \eqref{eq:big-system-unknowns-2} and \eqref{eq:big-system-unknowns-3} together amount to a linear system of 42 equations in 32 unknowns. Studying it in Mathematica, we find that it has a solution provided that 
\be 
 \sum_{i=1}^4 m_2^i \braket{ \ \dots \ } + \sum_{i=1}^4\sum_{j\neq i} \frac{x_i-x_j}{z_i-z_j} m_1^j \braket{m_1^j + \tfrac{1}{2}; m_2^i + \tfrac{1}{2}}  = 0 \ , 
\ee
which is indeed satisfied, as shown in the discussion below eq.~\eqref{eq:4pt-1111-rec-simple}. In fact the solution is unique and also satisfies 
\be 
A^\pm_1(z_4,x_4, u_2, u_1) = A^\pm_2(z_4,x_4, u_1, u_2) \ . 
\ee

\end{document}